\documentclass[5p]{elsarticle}

\usepackage{amsmath,amssymb,amsfonts,chemarrow,balance}
\usepackage{amsmath} 
\usepackage{graphicx}
\usepackage{subfigure}
\usepackage[hyperpageref]{backref}
\usepackage{makecell}
\usepackage{url}
\usepackage{mathenv}
\usepackage{color}
\usepackage{color}
\usepackage{rotating}
\usepackage{graphicx}
\usepackage{longtable}
\usepackage{supertabular,booktabs}
\usepackage{breqn}
\usepackage{tikz}
\usepackage{tikzpeople}
\usepackage{listings}
\usepackage{algpseudocode}
\newcommand{\BfPara}[1]{{\noindent\bf#1.}\xspace}
\usepackage{times}
\usepackage{xspace}
\usepackage{setspace}
\usepackage{multirow}
\usepackage{booktabs}

\usepackage[inline]{enumitem}

\definecolor{codegreen}{rgb}{0,0.6,0}
\definecolor{codegray}{rgb}{0.5,0.5,0.5}
\definecolor{codepurple}{rgb}{0.58,0,0.82}
\definecolor{backcolour}{rgb}{0.95,0.95,0.92}

\lstdefinestyle{mystyle}{
    backgroundcolor=\color{backcolour},   
    commentstyle=\color{codegreen},
    keywordstyle=\color{magenta},
    numberstyle=\tiny\color{codegray},
    stringstyle=\color{codepurple},
    basicstyle=\footnotesize,
    breakatwhitespace=false,         
    breaklines=true,                 
    captionpos=b,                    
    keepspaces=true,                 
    numbers=left,                    
    numbersep=5pt,                  
    showspaces=false,                
    showstringspaces=false,
    showtabs=false,                  
    tabsize=2
}

{\bfseries}{\itshape}
{\bfseries}{\itshape}
{\bfseries}{\itshape}
{\bfseries}{\itshape}
{\bfseries}{\itshape}
{\itshape}

\newcommand{\etal}{{\em et al.}\xspace}
\newcommand{\eg}{{\em e.g.},\xspace}
\newcommand{\note}[1]{}

\usepackage{pifont}
\newcommand{\cmark}{{\ding{51}}}
\newcommand{\xmark}{\ding{55}}%

\newcommand{\etc}{{etc.}\xspace}

\usepackage{float}

\usepackage{hyperref}
\definecolor{linkcolour}{rgb}{0,0.2,0.6}
\definecolor{xgreen}{rgb}{0.2,0.6,0.0}
\definecolor{xred}{rgb}{0.7,0.1,0.0}

\begin{document}

\begin{frontmatter}

\title{Domain Name System Security and Privacy: A Contemporary Survey}


\author[]{Aminollah Khormali}
\author[]{Jeman Park}
\author[]{Hisham Alasmary}
\author[]{Afsah Anwar}
\author[]{David Mohaisen}

\address{University of Central Florida}

\begin{abstract}

The domain name system (DNS) is one of the most important components of today's Internet, and is the standard naming convention between human-readable domain names and machine-routable IP addresses of Internet resources. However, due to the vulnerability of DNS to various threats, its security and functionality have been continuously challenged over the course of time. Although, researchers have addressed various aspects of the DNS in the literature, there are still many challenges yet to be addressed. In order to comprehensively understand the root causes of the vulnerabilities of DNS, it is mandatory to review the various activities in the research community on DNS landscape. To this end, this paper surveys more than 170 peer reviewed papers, which are published in both top conferences and journals in last ten years, and summarizes vulnerabilities in DNS and corresponding countermeasures. This paper not only focuses on the DNS threat landscape and existing challenges, but also discusses the utilized data analysis methods, which are frequently used to address DNS threat vulnerabilities. Furthermore, we looked into the DNS threat landscape from the view point of the involved entities in the DNS infrastructure in an attempt to point out more vulnerable entities in the system. 


\end{abstract}

\end{frontmatter}


\section{Introduction}

The Domain Name System (DNS) is one of the pillars of the operation of the Internet, the medium on which most of communications are today transported. Among other purposes, DNS is used today widely for domain names to translate them into Internet Protocol (IP) addresses. The domain name system maps a name that people use to locate a website to the IP address that a computer uses to locate that website. For example, if a user types www.example.com into a web browser, a server behind the scenes will map that name to an IP address, e.g., {1.2.3.4}, corresponding to that domain. Web browsing and most other Internet activities, e.g., transferring files, rely on DNS to quickly provide the information necessary to connect users to remote hosts. DNS mapping is distributed throughout the Internet in a hierarchy of authorities. 

Based on the key role of the DNS in networking infrastructure, attackers are aggressively looking for new ways to compromise the DNS infrastructure. Therefore, it is necessary to understand the evolution of DNS security and the associated issues to preserve reliable and secure services, and to improve the security of DNS through various iterations of design revisions. Indeed, recently, researchers from academia and industry have focused on improving the design of DNS by allowing various necessary options, and by revising its assumptions of operation. However, the constant evolution of DNS, as well as the rise of new detrimental and unconventional issues, such as pervasive adversaries, privacy risks, and new and advanced forms of attacks, make the functionality, security, and privacy of DNS important issues that require continued attention and investigation. To this end, surveying, summarizing, and categorizing the body of work on DNS security and privacy is an  important endeavor that is lacking in the existing literature. This effort is necessary to guide the community to what is an open problem that requires further attention. 

Particularly, despite the large number of works in the literature on DNS security and privacy, existing contributions are scattered across different research areas and a comprehensive yet concise survey is lacking. Although there exist several surveys in the field of DNS~\cite{ZouZPPLL16,KangSM16,TorabiBAD18,ZhauniarovichKYD18,SpauldingUM16}, they are limited from several points of view. First, the scope of those works is narrow, e.g., covering only a certain aspect of the DNS ecosystem, such as detecting Internet abuse by analyzing passive DNS traffic~\cite{TorabiBAD18}, malicious domain detection through DNS data analysis~\cite{ZhauniarovichKYD18}, and investigating the domain name squatting ecosystem~\cite{SpauldingUM16}. Second, the number of studies covered in each of those prior works---even on the very well accepted set of problems in the field of DNS security and privacy---is limited, and does not cover the most recent advances, which require further attention and consideration---especially in an evolving domain~\cite{ZouZPPLL16,SpauldingUM16}. 

\begin{figure*}[t]
\begin{center}
  \includegraphics[width=0.9\textwidth]{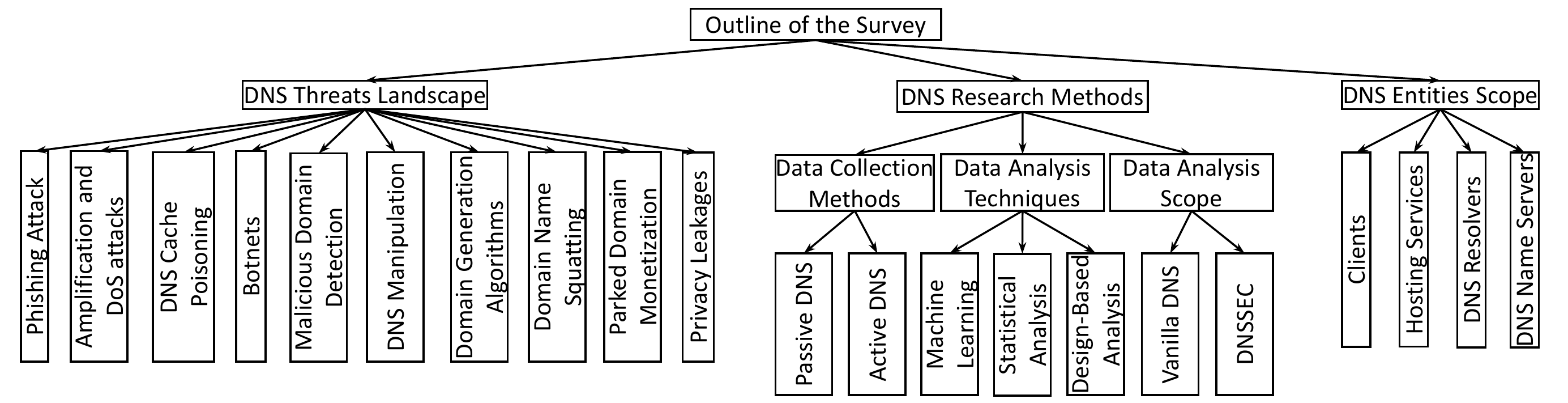}
  \caption{The general outline of the survey. We have surveyed a wide variety of research works from various points of view, including DNS threat landscape, DNS research methods, and DNS research scops. }
  \label{fig:SurveyOutline}  
\end{center}
\end{figure*}

To this end, this survey fills the gap in the literature by putting together an up-to-date summary of research works that cover various aspects of the DNS ecosystem and associated security and privacy issues---and all in one place. Toward this goal, we have surveyed a wide range of research works from various points of view, including the DNS threat landscape, DNS research methods, and DNS research scopes. 

The general outline of the survey, highlighting the road map towards the key terms and directions, is shown in \autoref{fig:SurveyOutline}. In the following we briefly review some of those broad directions.

\subsection{DNS Threat Landscape} A central component of this work is DNS threat landscape. By reviewing the threats, we primarily focus on the common challenges in security and privacy of DNS infrastructure. Such threats may not only compromise the security and privacy of end-users, but may also undermine the performance and availability of Internet services that heavily rely on DNS for their operation. We demonstrate such an interdependence through various practical and challenging attacks and efforts in the literature to defend against them. For example, attackers abuse the functionality of open DNS resolvers to transfer a small number of DNS queries into significantly large payload coordinated at a target server or network, thus making the server inaccessible, in what is known as amplification attacks. Despite the large body of research works on detection and mitigation of \textit{DNS amplification attacks}~\cite{AizuddinANNAA17, ballaniF16, KimLCAJK17, HerzbergS14, MacFarlandSK15, PerdisciCG12, TruongC16, VermaHHHRKF16, zheng2018realtime}, such attacks still compromise today's Internet, according to \textit{Nexusguard's} 2018 threat report of Q1~\cite{Nexusguard18}. 

Another example we highlight is cache poisoning; DNS resolver cache can be compromised through the insertion of unauthorized domain names and IP addresses in order to redirect users to malicious IP addresses. Consequently, by launching this attack an adversary can compromise the performance of Internet services and distribute malicious contents, such as malware, spam, credentials theft, phishing contents, among many others. There exist several research works that attempted to address \textit{DNS cache poisoning attacks}~\cite{ShulmanW14, AntonakakisDLPLB10, ChenMP15, PerdisciALL09, WuDZW15, ChenAPNDL14, HaoW17} using different methods, such as DNSSEC~\cite{HerzbergS12, ShulmanW14}, WSEC DNS~\cite{PerdisciALL09}, ECO-DNS~\cite{ChenMP15}, and Anax~\cite{AntonakakisDLPLB10}, among others, which we outline this study.

Attackers also utilize \textit{botnets} that allow them to access devices through propagation and coordinate among them using C\&C channels to perform wide variety of malicious activities, such as Distributed Denial of Service (DDoS) attack~\cite{rustock}, amplification attacks, and credentials stealing,  among many others~\cite{riccardi2013titans}. DNS plays an essential role in the operation of botnets by providing easily deploy-able C\&C channel. Although various research works have studied botnets and possible solutions to defend against their attacks~\cite{jinYGY2015, JiangLLLDW12, TruongC16, PlohmannYKBG16, XuBSY13}, including marginalizing DNS infrastructure used for this attack. However, such attacks are still on the rise, compromising users' security. Specifically, the emergence and rise of IoT and associated vulnerabilities opened new directions, whereby attack vectors facilitated by DNS' reliability and deployment played a crucial role in those attacks. For example, the Mirai botnet attacks  IoT devices, such as routers and video recorders~\cite{whittaker_2016, palmer_2018}, and uses them for amplification attacks, necessitating  security solutions optimized for IoT and its interplay with DNS~\cite{kolias2017ddos, bertino2017botnets}. 

\textit{DNS manipulation} may expose users to multiple threats, such as phishing and malicious domains. Accordingly, a large number of studies considered DNS manipulation~\cite{WeaverKNP11, SchompCRA14, KuhrerHBRH15, PearceJLEFWP17, TrevisanDMM17}, and its interplay with the detection and mitigation of \textit{phishing attacks}~\cite{BilgeKKB11, DagonPLL08, KintisMLCGPNA17, CookGD08,MedvetKK08, HaraYM09}. Despite this large number of studies, phishing and associated DNS manipulation is still a challenging and expanding area.  Phishers are targeting mobile messaging and social apps at an increasing rate~\cite{wandera_2018}, highlighting the need for further investigation and defenses. 

A large body of research has been devoted to the detection of \textit{malicious domain names}~\cite{AntonakakisPDLF10, AntonakakisPLVD11, BilgeKKB11, BilgeSBKK14, CanaliCVK11, FelegyhaziKP10, GaoYCPGJD13,HaoKMPF16, HaoTPFKGH13,jinYGY2015, KhalilGNY18, LuoTZSLNM15, MaSSV09,HaoFP11, LiuLDWLD17}, which can be classified into two general categories: classification based~\cite{AntonakakisPDLF10,AntonakakisPLVD11,BilgeKKB11,BilgeSBKK14,jinYGY2015} and inference based~\cite{KhalilGNY18,GaoYCPGJD13,LuoTZSLNM15,LiuLDWLD17} methods. Classification-based methods rely on local network and host information. The inference-based methods exploit global information along with local information of domains. 

Domain Generation Algorithms (DGAs) are utilized to intermittently produce a large number of domain names for malicious purposes. There exist several research works that explored the ecosystem of DGAs~\cite{YadavRRR10, AntonakakisPNVALD12, baraboschWLG12, PlohmannYKBG16, TruongC16, SpauldingPKM18}. 

Furthermore, another type of threat is \textit{domain name squatting}, where cybersquatters register variants of popular trademark names through different squatting strategies, has been considered. Such attacks include various variants, i.e. typosquatting~\cite{DWangBWVD06,BanerjeeBFB08, SpauldingUM16, AgtenJPN15, SzurdiKCSFK14, KhanHLK15, SpauldingNM17, SpauldingUM17}, bitsquatting~\cite{Dinaburg, NikiforakisAMDPJ13, VissersBGJN17}, combosquatting~\cite{KintisMLCGPNA17}, and soundsquatting~\cite{NikiforakisBDPJ14}. Domain name squatting becomes popular as an attack vector to start many other types of attacks~\cite{SzurdiKCSFK14,AgtenJPN15}. Related to this is domain monetization through parking; despite technically being a legitimate business, \textit{parked domain monetization} can often lead to and get mixed with other practices, i.e. click fraud, as well as traffic spam and traffic stealing---which usually lead to serving  malicious content i.e. malware and  spam~\cite{AlrwaisYALW14, SzurdiKCSFK14, WeaverKP11, VissersJN15}. All of these threats may compromise the security and privacy of DNS landscape and users. There are large amount of research regarding \textit{privacy leakage} and various solutions have been proposed, such as  EncDNS~\cite{HerrmannFLF14}, PPDNS~\cite{LuT10}, DLV~\cite{MohaisenGR17}, \etc However,

\BfPara{DNS Research Methods} Another vertical of our survey is the DNS research methods used in the literature, and their associated work. In exploring the DNS research methods, we have investigated various data analysis methods, which have been utilized to detect, model, and mitigate the aforementioned DNS threats. First, we describe two main \textit{DNS data collection methods} utilized in the literature and the associated works, including \textit{passive DNS data (PDNS)}~\cite{AlrwaisLMWWQBM17, AntonakakisPLVD11, CallahanAR13, Tajalizadehkhoob17a, KountourasKLCND16, TruongC16, PerdisciCG12, LiAXYW13, BilgeSBKK14, KintisMLCGPNA17,  KhanHLK15, HaoW17, GaoYCPGJD13} and \textit{Active DNS data (ADNS)}~\cite{ KhanHLK15, KountourasKLCND16, VanJSP16, KhalilGNY18, KintisMLCGPNA17, ChungR0CLMMW17}.  Next, we categorize the research works based on the common \textit{DNS data analysis techniques} that have been used in the literature, such as\textit{ machine learning} algorithms~\cite{AlrwaisLMWWQBM17, AntonakakisDLPLB10, AntonakakisPDLF10, HaoTPFKGH13, HaoW17, KhalilGNY18, LuoTZSLNM15, MaSSV09, RadwanH17, RahbariniaPA15, ShulmanW15, VissersJN15} and \textit{association analysis}~\cite{YadavRRR10, Tajalizadehkhoob17a, HockK16, KhalilGNY18, GomezNA17}. Finally, we categorize the research works based on their scope of analysis and from security point of view into \textit{vanilla DNS}~\cite{SchompCRA14, DagonPLL08, XuBSY13, JacksonBBSB09, SchompAR14, DagonAVJL08, ChenMP15} and \textit{DNSSEC}~\cite{AdrichemBLWWFK15, BabuP18,BauM10, ChungR0CLMMW17, ChungRCLMMW17, HerzbergS13, HerzbergS12, HerzbergS14, JalalzaiSI15}.

\subsection{DNS Entities Scopes} We categorize the literature based on the entities they target, including \textit{DNS name servers}, \textit{open DNS resolvers}, \textit{hosting providers}, and \textit{clients}. Such categorization would reveal the most susceptible entity and its impact on the performance and security of whole DNS ecosystem. For example, open DNS resolvers can be exploited to conduct multiple attacks on behalf of attackers~\cite{AgerMSU10, AntonakakisDLPLB10, ballaniF16, GaoYCPGJD13, HerrmannFLF14, KuhrerHBRH15, HerzbergS12, PearceJLEFWP17, PerdisciALL09, SchompAR14, SchompCRA14, VermaHHHRKF16}.

We discuss advantages and disadvantages of the proposed countermeasures at the end of each section and highlighted the existing open problems. We list the abbreviations used throughout this paper in~\autoref{tab:abbreviations}. Finally, the key characteristics of our survey in this paper can be summarized as: 
\begin{enumerate*}
\item we conduct a comprehensive survey mainly focusing on recent major advances in the area of DNS security and privacy. The focuses of the survey are the threat landscape, research methods, and research scopes (based on a classification of the DNS system model).

\item The scope of the survey is defined in such a way that not only it focuses on the up-to-date challenges, by gathering works mostly published in the past decade, but also by focusing on the most critical and ongoing challenges corresponding to the current trend.  

\item We highlight the existing open challenges in the security and privacy of DNS infrastructure, which requires further investigations by security specialists to be addressed. We use that as a summary in every surveyed area.
\end{enumerate*}


\subsection{Organization} In section~\ref{sec:Preliminaries} we present an overview of DNS functionality and DNS resolution. Major DNS threats and associated countermeasures are presented in section~\ref{sec:Threats}. DNS research methods are presented in section~\ref{sec:Methods}. In section~\ref{sec:Entities} we explore research work addressing the different entities in the DNS model. Finally, the paper is concluded in section~\ref{sec:Conclusion}.

\begin{table*}[t]
\begin{center}
\caption{List of abbreviations in alphabetical order.}
\label{tab:abbreviations}
\scalebox{0.99}{
\begin{tabular}{ll|ll}
\Xhline{2\arrayrulewidth}
\textbf{Term}  & \textbf{Definition} & \textbf{Term}  & \textbf{Definition}   \\
\Xhline{2\arrayrulewidth}
%
2LD & Second Level Domain	&	ML	&	Machine	Learning\\		
3LD & Third Level Domain	&	MLP	&	Multi	Layer	Perceptron\\	
AP & Affinity Propagation  	&	MS	&	Mean	Shift	\\	
BGP & Border Gateway Protocol 	&	NAT	&	Network	Address	Translation\\	
C\&C & Command and Control	&	NB	&	Naive	Bayes\\		
CPM & Convex Polytope Machine 	&	NN		&	Neural	Networks	\\
DDoS & Distributed Denial of Service	&	RBC	&	Rule-Based	Classifier	\\	
DGA & Domain Generation Algorithm 	&	RDNS	& Recursive	Domain Name System	\\
DNS & Domain Name System	&	RF	&	Random	Forest\\		
DNSSEC & DNS Security Extensions	&	RR	&	Resource	Records\\			
DoS & Denial of Service	&	RSA	&	Rivest	Shamir	Adleman	\\	
DT & Decision Tree	&	SDN	&	Software	Defined	Networking\\		
HTTP & Hypertext Transfer Protocol 	&	SL	&	Simple	Logistic\\			
IoT & Internet of Things	&	SVM	&	Support	Vector	Machine\\		
IP & Internet Protocol 	&	TCP	&	Transmission Control	Protocol\\		
IPv4 & Internet Protocol Version 4	&	TLD	&	Top	Level Domain\\		
IPv6 & Internet Protocol Version 6	&	TLS	&	Transport	Layer	Security	\\	
ISP & Internet Service Provider	&	TSVM	&	Transductive	Support	Vector	Machine	\\
KNN & K-Nearest Neighbour 	&	TTL	&	Time	To	Live\\		
LAN & Local Area Network 	&	UDP	&	User	Datagram	Protocol\\		
LWL & Locally Weighted Learning 	&	URL	&	Uniform	Resource	Locator\\
MITM & Man in the Middle	&	& \\ 
\Xhline{2\arrayrulewidth}
\end{tabular}
}
\end{center}
\end{table*}

\section{Preliminaries}\label{sec:Preliminaries}
This section is devoted to provide an overview of how DNS works, including the domain name resolution process. 

\subsection{Functionality}

When a user wants to visit a website through the Internet he does so by connecting to a domain name, such as www.example.com; however, computers do not communicate with domain names, but rather they are only able to communicate using addresses represented as numbers, namely IP addresses. Therefore, the domain name of the intended website should be converted to its associated IP address. Addresses in the network can be represented either using IPv4 or IPv6. The first is composed of four bytes and is represented using four number segments, such as 1.2.3.4, whereas IPv6 is four times larger and is composed of eight segments, each of them is two bytes and represented by hexadecimal numbers.  In comparison to natural languages, recognizing and remembering numerical addresses of the intended Internet services is not an easy task for users. Therefore, in the early 1980's Paul Mockapetris \cite{rfc883} introduced the basics of the DNS, which enables users to automatically map human-readable domain names into machine-readable addresses. DNS is an essential component of the functionality of the Internet. 

\subsection{DNS Resolution}

\begin{figure}[t]
\begin{center}
\begin{tikzpicture}

\node[bob, monitor, minimum size=0.7cm,xshift=-2.5cm, yshift=2cm]{\sf client};
\node[builder, monitor, mirrored, minimum size=0.7cm,xshift=1.2cm, yshift=2cm]{};
\node at (0.85,1.3) [black] {\sf recursive};
\node[businessman, monitor, mirrored, minimum size=0.7cm,xshift=5.2cm, yshift=4.0cm]{\sf root};
\node[businessman,monitor, mirrored, minimum size=0.7cm,xshift=5.2cm, yshift=2.0cm]{\sf TLD};
\node[businessman,monitor, mirrored, minimum size=0.7cm,xshift=5.2cm]{};
\node at (4.7,-0.7) [black] {\sf authoritative};
\draw[->,black,thick] (-1.7,1.9) -- (0.4,1.9);
\node at (-0.75,2.3) [black] {{\footnotesize \textcircled{\tiny 1}} {\sf \footnotesize www.example.com?}};
\node at (-0.7,1.45) [black] {{\footnotesize \textcircled{\tiny 8}} {\sf \footnotesize IP=1.2.3.4}};
\draw[->,black,thick] (0.4,1.7) -- (-1.7,1.7);

\draw[->,black,thick] (1.6,2.6) -- (4.4,4.0);
\draw[->,black,thick] (4.4,3.8) -- (1.6,2.4);
\node at (2.9,3.5) [rotate=25, black] {{\footnotesize \textcircled{\tiny 2}} {\sf \footnotesize www.example.com?}};
\node at (3.2,2.95) [rotate=25, black] {{\footnotesize \textcircled{\tiny 3}} {\sf \footnotesize .com}};

\draw[->,black,thick] (1.6,2.0) -- (4.4,2.0);
\draw[->,black,thick] (4.4,1.85) -- (1.6,1.85);
\node at (3.1, 2.2) [black] {{\footnotesize \textcircled{\tiny 4}} {\sf \footnotesize www.example.com?}};
\node at (3.1, 1.7) [black] {{\footnotesize \textcircled{\tiny 5}} {\sf \footnotesize example.com}};

\draw[->,black,thick] (1.6,1.6) -- (4.4,-0.1);
\draw[->,black,thick] (4.4,-0.3) -- (1.6,1.4);
\node at (3.25, 0.8) [rotate = 329, black] {{\footnotesize \textcircled{\tiny 6}} {\sf \footnotesize www.example.com?}};
\node at (3.05, 0.32) [rotate = 329, black] {{\footnotesize \textcircled{\tiny 7}} {\sf \footnotesize IP=1.2.3.4}};

\end{tikzpicture}
\caption{Illustration of DNS resolution over recursive, root, TLD, and authoritative name server. Once a user attempts to access www.example.com (step \textcircled{\tiny 1}), in the case that there is no information about the given domain in the local cache and host table, recursive resolver starts asking root, TLD, and authoritative name servers and over steps \textcircled{\tiny 2} through \textcircled{\tiny 7} to answer the query.}

\label{fig:DNSResolution}
\end{center}
\end{figure}
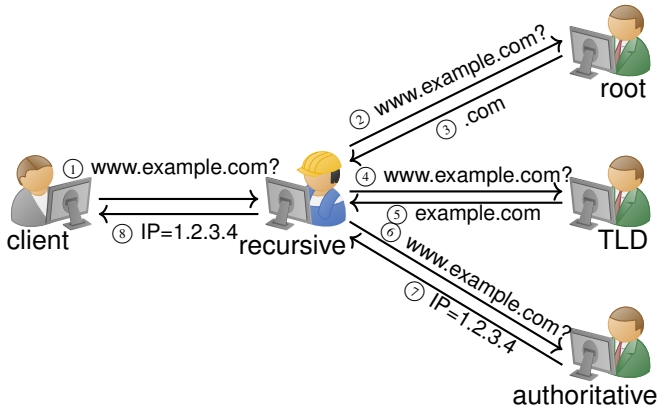

DNS is a hierarchical and globally distributed directory service that utilizes root, TLD, and authoritative name servers in order to resolve string addresses of the domains into IP addresses. The process of domain name resolution is shown in \autoref{fig:DNSResolution}.The DNS resolution process begins once a user attempts to access a web service using Internet web browser. In the case where there is no information about the given domain in the local cache and host table, the local resolver initiates a DNS query to a recursive resolver seeking to match the domain with its corresponding IP address. Then, the recursive resolver starts asking the root, TLD, and then authoritative name servers, and over steps \textcircled{\tiny 2} through \textcircled{\tiny 7}, to answer the query. The root server is the first server that receives queries from recursive resolvers, in step \textcircled{\tiny 2}. The root servers are globally distributed and maintain the IP addresses and location of authoritative TLD name servers. In step \textcircled{\tiny 3}, the root name server replies to the query with the appropriate list of authoritative TLD servers for the .com TLD. In step \textcircled{\tiny 4}, the recursive server sends a query for example.com to the .com TLD name server. Once the query reaches the .com TLD name server, it responds with the IP address of the domain's authoritative name server, as shown in step \textcircled{\tiny 5}. At the next step, namely step \textcircled{\tiny 6}, the recursive resolver sends a query to the authoritative name server. The authoritative name server knows the IP address for www.example.com and at step \textcircled{\tiny 7} that answer is returned to the recursive name server. Finally, the determined IP address of the requested domain name is forwarded to the local resolver and then to a web browser. As a result, the browser can send a Hypertext Transfer Protocol (HTTP) request to the website to retrieve its contents. 

\section{DNS Threat Landscape} \label{sec:Threats}

The original implementation of the DNS did not include security and privacy protections; instead the original DNS was built to be a scalable distributed system. As the Internet has grown, however, adversaries have found weaknesses that can be abused for malicious purposes. There is a large body of research in this area, which lead to identification of major threats that substantially compromise the DNS security. In this section, we would delve into the common threats in the DNS ecosystem and state-of-the-art countermeasures presented in the literature. 

Our effort in enumerating and categorizing the literature in this domain is featured in the follow. We review the proposed methods for detection and mitigation of popular threats of DNS infrastructure, including amplification and DoS attacks~(\textsection\ref{sec:Amplification}), DNS cache poisoning attacks~(\textsection\ref{sec:CachePoisoning}), botnets attacks~(\textsection\ref{sec:Botnet}), phishing attacks~(\textsection\ref{sec:Phishing}), DNS manipulation~(\textsection\ref{sec:Manipulation}), malicious domain detection~(\textsection\ref{sec:MaliciousDomain}), domain generation algorithms~(\textsection\ref{sec:DGA}), domain name squatting~(\textsection\ref{sec:Squatting}), parked domain monetization and associated security issues~(\textsection\ref{sec:Monetization}), and privacy leakage~(\textsection\ref{sec:Privacy}). 

In addition, we will discuss challenges and highlight open problems that require further investigations by security researchers associated with each of those major problems. In~\autoref{tab:ThreatsCont.} and~\autoref{tab:ThreatsCont2.}, topical classification of DNS threats in the literature is addressed, with sample works presented as well.

\subsection{Amplification and DoS Attacks}\label{sec:Amplification}

\begin{figure}[t]
\begin{center}
\begin{tikzpicture}

\node at (-0.4,5.3) [rotate = 0, black] {\sf botmaster};
\node[devil, monitor, minimum size=0.7cm,xshift=-2.1cm, yshift=5.2cm]{};

\node at (2,5) [rotate = 0, black] { \textcircled{\tiny 2} {\sf \footnotesize command bots}};

\node at (5.25,3.5) [rotate = 90, black] {\sf bots};
\node[criminal, monitor, mirrored, minimum size=0.7cm,xshift=0cm, yshift=3.6cm]{};
\node[criminal, monitor, mirrored, minimum size=0.7cm,xshift=1.2cm, yshift=3.6cm]{};
\node[criminal, monitor, mirrored, minimum size=0.7cm,xshift=2.4cm, yshift=3.6cm]{};
\node[criminal, monitor, mirrored, minimum size=0.7cm,xshift=4.7cm, yshift=3.6cm]{};

\node at (5.25,-0.75) [rotate = 90, black] {\sf open RDNS};
\node[builder, monitor, mirrored, minimum size=0.7cm,xshift=0cm, yshift=-0.75cm]{};
\node[builder, monitor, mirrored, minimum size=0.7cm,xshift=1.2cm, yshift=-0.75cm]{};
\node[builder, monitor, mirrored, minimum size=0.7cm,xshift=2.4cm, yshift=-0.75cm]{};
\node[builder, monitor, mirrored, minimum size=0.7cm,xshift=4.7cm, yshift=-0.75cm]{};

\node at (-1.1,0.18) [rotate = 90, black] { \textcircled{\tiny 4} {\sf \footnotesize request}};
\node at (-1.1,-1.9) [rotate = 270, black] { \textcircled{\tiny 5} {\sf \footnotesize response}};

\node at (-2,-1.6) [rotate = 0, black] {\sf zone file};
\node[businessman, monitor, minimum size=0.7cm,xshift=-2.3cm, yshift=-0.75cm]{};

\node at (4.2,-2.1) [rotate = 0, black] { \textcircled{\tiny 6} {\sf \footnotesize response}};

\node at (3,-2.85) [rotate = 0, black] {\sf victim};
\node[bob, monitor, mirrored, minimum size=0.7cm,xshift=2cm, yshift=-2.6cm]{};

\draw [-stealth,black,thick](1.25,3) -- (1.25,-0.15);
\draw [-stealth,black,thick](2.5,3) -- (2.5,-0.15);
\draw [-stealth,black,thick](0,3) -- (0,-0.15);
\draw [-stealth,black,thick](4.7,3) -- (4.7,-0.15);

\draw [-stealth,black,thick] (-1.3,5.1) -- (4,4.2);
\draw [-stealth,black,thick](-1.3,5.1) -- (2.5,4.2);
\draw [-stealth,black,thick](-1.3,5.1) -- (0,4.2);
\draw [-stealth,black,thick](-1.3,5.1) -- (1.25,4.2);

\draw [-stealth,black,thick] (-2,4.5) -- (-2,-0.2);

\draw [-stealth,red,ultra thick](0,-1.35) -- (1.45,-2.2);
\draw [-stealth,red,ultra thick](4.3,-1.35) -- (2.4,-2.25);
\draw [-stealth,red,ultra thick](2.5,-1.35) -- (2.2,-2);
\draw [-stealth,red,ultra thick](1.25,-1.35) -- (1.7,-2);

\draw [-stealth,black,thick](-0.8,-0.65) -- (-1.5,-0.65) ;
\draw [-stealth,black,thick](-1.5,-1)  -- (-0.8,-1) ;

\node at (3.4,3.4) [rotate =0 , black] {\Large {\Large \textbf{- - -}}};
\node at (3.4,-1) [rotate =0 , black] {\Large {\Large \textbf{- - -}}};

\node at (-2.3,2.2) [rotate = -270, black] { \textcircled{\tiny 1} {\sf \footnotesize corrupt authoritative DNS server}};

\node at (-0.3,1.5) [rotate = -270, black] { \textcircled{\tiny 3} {\sf \footnotesize spoofed DNS query}};
\node at (1,1.5) [rotate = -270, black] { \textcircled{\tiny 3} {\sf \footnotesize spoofed DNS query}};
\node at (2.3,1.5) [rotate = -270, black] { \textcircled{\tiny 3} {\sf \footnotesize spoofed DNS query}};
\node at (4.45,1.5) [rotate = -270, black] { \textcircled{\tiny 3} {\sf \footnotesize spoofed DNS query}};

\end{tikzpicture}
\caption{High-level architecture of a typical DNS amplification attack. Small number of DNS queries are being transformed into a significantly large payload coordinated at the target network.}

\label{DNSAmplification}
\end{center}
\end{figure}
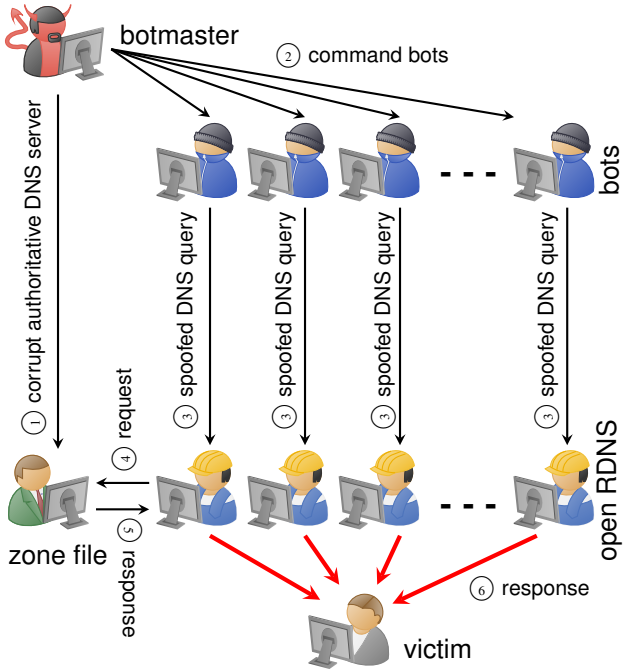

A DNS amplification attack is a reflection-based DDoS attack. In DNS, an amplification attack is done by issuing a small number of DNS queries that are later  transformed into a considerably large payload coordinated at the target network. The high level architecture of a typical DNS amplification attack is demonstrated in~\autoref{DNSAmplification}. The attacker hides the exploit source and directs the DNS response into the target address through spoofing look-up requests issued to DNS servers. It is difficult to defend against such an attack, since it originates from legitimate servers with legitimate traffic. 

A wide variety research works have been conducted to detect and mitigate DNS amplification attacks~\cite{ballaniF16, HerzbergS14, Rossow14, RijswijkDeijSP14, MacFarlandSK15, VermaHHHRKF16, AizuddinANNAA17, PerdisciCG12, TruongC16, KimLCAJK17, zheng2018realtime}. For instance, Ballani \etal~\cite{ballaniF16} have presented a simple method based on caching behavior analysis of DNS resolvers to defend against DNS DoS attacks. They have stored cached records with TTL in a stale cache. Then, the stale cache can be used by a resolver that does not receive any response from the authoritative name servers.
Herzberg \etal~\cite{HerzbergS14} have designed an anti-reflection system, providing DNS authentication, which nullifies the amplification factor of the DNS responses abused for DoS attacks. DNS authentication is composed of two subsystems, namely a {\em request authentication} that detects and filters requests sent from spoofed IP addresses, and {\em resolver authentication} that maintains a list of potentially compromised hosts. They have deployed the resolver authentication as a cloud-based service to further reduce costs and provide additional defenses for DNS servers. 
Furthermore, Rijswijk-Deij \etal~\cite{RijswijkDeijSP14} investigated the potential for abuse in DNSSEC-signed domains in a large scale, covering 70\% of all signed domains in operation. Their analysis demonstrate that DNSSEC in-fact empowers DNS amplification attacks for a particular query type, \textit{ANY}. In addition, Rossow~\cite{Rossow14} has investigated Distributed Reflective Denial-of-Service (DRDoS) attacks through revisiting well-known UDP-based protocols, including network services, online games, file sharing networks, and botnets to assess their security against DRDoS abuse. His analysis revealed that attackers already started abusing 14 protocols through bandwidth amplification and multiplying the traffic up to a factor 4670.

\begin{table}[t]
\begin{center}
\caption{Summary of representative work that studied DNS amplification attacks; RL is Response Latency, ID is Input Data, and O{$^{(1,2,3)}$} donates Other disadvantages: {$^{(1)}$}modification of standard DNS resolver semantics, {$^{(2)}$}third-party organizations, and {$^{(3)}$}low detection accuracy.}
\label{tab:SumAmplification}
\scalebox{0.7}{
\begin{tabular}{l|l|l|c|c|c}
\Xhline{2\arrayrulewidth}
\multirow{2}{*}{\textbf{Work}} & \multirow{2}{*} {\textbf{Method}} & \multirow{2}{*}{\textbf{Advantages} } & \multicolumn{3}{c}{\textbf{Disadvantages}} \\
\cline{4-6}
                            &  & & RL & ID & O{$^{(1,2,3)}$} \\
                            \Xhline{2\arrayrulewidth}
\cite{AizuddinANNAA17}    & sFlow/SDN    & Near real-time detection         & \xmark & &   \\ \hline
\cite{ballaniF16}          & Stale cache                        & Simple, lightweight           & & &  \xmark{$^{(1)}$}  \\ \hline
\cite{KimLCAJK17}              & One-to-one mapping          & Complete mitigation              & \xmark &  &\\ \hline
\cite{HerzbergS14}        & DNS authentication                 & Cloud based service              & \xmark & &             \\ \hline
\cite{MacFarlandSK15}   & Tunnel/remote resolver          & Legacy compatible    & \xmark & & \xmark{$^{(2)}$}  \\ \hline
\cite{TruongC16}            & ML/domain features     & Small feature space   & & & \xmark{$^{(3)}$} \\ \hline
\cite{VermaHHHRKF16}         & DRS-ADAM                           & Easy deployment                 & & \xmark &    \\ \hline
\cite{zheng2018realtime}     & RADAR                              & Real-time detection              & & \xmark & \\
\Xhline{2\arrayrulewidth}
\end{tabular}}
\end{center}
\end{table}

MacFarland \etal~\cite{MacFarlandSK15} examined a large number of domains ($\approx$129M) and authoritative servers ($\approx$1.1M) to investigate the inherent DNS amplification risks associated with DNS authoritative name servers. Their analysis showed that only a small number of authoritative servers ($\approx$3.8\%) are responsible for the highest amplification factors. In addition, their analysis revealed that adoption of DNS response rate limit is limited to less than $\approx$3\% of authoritative servers. Finally, they have suggested tunnelling into a remote resolver as a straightforward and simple countermeasure to mitigate on-going attack at the organization level. 
Verma \etal~\cite{VermaHHHRKF16} have utilized the fact that DNS resolvers share the local DNS query rates to propose an amplified DNS attack mitigation system called Distributed Rate Sharing-based Amplified DNS-DDoS Attack Mitigation (DRS-ADAM). The authors claim that DRS-ADAM detects and \textit{completely stops} an amplified DNS attack by imposing DNS query rate sharing among resolvers involved in an attack. DRS-ADAM has several advantages; deployment, robustness against manipulation, and attack mitigation.  

Aizuddin \etal~\cite{AizuddinANNAA17} have incorporated sFlow with security-centric SDN features to analyze DNS query identifiers (IDs) for detecting and mitigating DNS amplification attack in a timely manner. Their analysis showed that the proposed method provides accurate detection results (more than 97.0\%) even with a small number of flow values (DNS attributes). 
Kim \etal~\cite{KimLCAJK17} have presented a DNS amplification attack mitigation system through a one-to-one strict mapping method between DNS requests and responses in order to identify {\em orphan DNS responses}. Their analysis showed that the proposed solution removes the possibility of false positive packets. 
Zheng \etal~\cite{zheng2018realtime} have proposed Reinforcing Anti-DDoS Actions in Real-time (RADAR) which detects various DDoS attacks, such as link flooding, SYN flooding, and UDP-based amplification attacks. They do so through adaptive correlation analysis on commercial off-the-shelf SDN switches. RADAR does not require any changes in the SDN protocols and switches deployed in the network today, nor does it require additional appliances to detect attacks, making it an easy-to-plug solution in today's operations. 

Truong \etal~\cite{TruongC16} analyzed DNS traffic to design a detection system for recognition of pseudo-random domain names, including Conficker and Zeus, form legitimate domain names. The proposed detection system is composed of two main subsystems, including feature extraction and classification. The length of domain names and their expected values construct the feature space and classification section is composed of several classification algorithms, e.g., RF, KNN, SVM, and NB.

\BfPara{Discussion} 
Despite the large body of research work on the detection and mitigation of DNS amplification and DoS attacks, such attacks are still prevalent and compromising today's Internet. \autoref{tab:SumAmplification} summarizes the proposed methods in the literature and their strengths and weaknesses. Shortcomings of the proposed methods, which would require further attention from the community, can be summarized as: 
\begin{enumerate*}
    \item increasing response latency due to detection process, which requires light-weight and latency sensitive detectors~\cite{HerzbergS14,MacFarlandSK15,AizuddinANNAA17,KimLCAJK17}.
    \item requiring a range of changes to the DNS resolvers and semantics~\cite{ballaniF16}, which calls for work that address legacy-compatibility or require very little of such changes. 
    \item requiring large number of IP addresses to be collected every day~\cite{PerdisciCG12}, or requiring large number of flow rules~\cite{zheng2018realtime}, which calls for aggregate and light-weight feature engineering methods. 
    \item low detection accuracy, in some cases~\cite{TruongC16}, which calls for improving accuracy through multi-modality of detection features. 
\end{enumerate*}

\subsection{DNS Cache Poisoning}\label{sec:CachePoisoning}

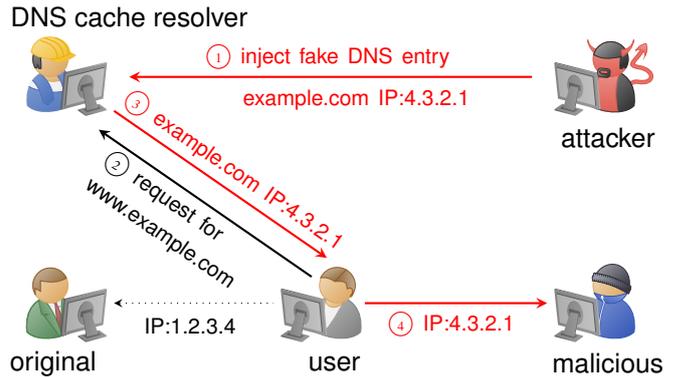
\begin{figure}[t]
\begin{center}
\begin{tikzpicture}

\node[devil,monitor, mirrored, minimum size=0.7cm, xshift=4.8cm, yshift=0cm]{};
\node[xshift=4.8cm, yshift=-0.8cm]{\sf attacker};

\node[ xshift=3cm, yshift=0.25cm, rotate = 0,  text width=7cm, node distance=1cm, red]{\textcircled{\tiny 1} {\sf \footnotesize inject fake DNS entry}};
\node[ xshift=3.5cm, yshift=-0.3cm, rotate = 0,  text width=7cm, node distance=1cm, red]{ {\sf \footnotesize example.com IP:4.3.2.1}};
\draw [stealth-,red,thick] (-1.5,-0) -- (3.8,0);

\node[builder, monitor, minimum size=0.7cm,xshift=-2.5cm, yshift=0cm]{};
\node[xshift=-1.5cm, yshift=0.8cm]{\sf DNS cache resolver};

\node[ xshift=0.5cm, yshift=-1.7cm, rotate = -35,  text width=5cm, node distance=1cm, red]{\textcircled{\tiny 3} {\sf \footnotesize example.com IP:4.3.2.1}};
\draw [-stealth,red,thick] (-1.7,-0.45) -- (1.1,-2.4);
\node[ xshift=-0.7cm, yshift=-2.1cm, rotate = -35,  text width=3cm, node distance=1cm, ,black]{\textcircled{\tiny 2} {\sf \footnotesize request for www.example.com}};
\draw [stealth-,black,thick] (-1.9,-0.7) -- (0.9,-2.65);

\node[businessman, monitor, minimum size=0.7cm,xshift=-2.5cm, yshift=-3cm]{};
\node[xshift=-2.5cm, yshift=-3.8cm]{\sf original};

\node[ xshift=0.2cm, yshift=-3.3cm, rotate = 0,  text width=3cm, node distance=1cm, ,black]{ {\sf \footnotesize IP:1.2.3.4}};
\draw [stealth-,black,dotted] (-1.7,-3) -- (0.4,-3);

\node[criminal,monitor, mirrored, minimum size=0.7cm, xshift=4.8cm, yshift=-3cm]{};
\node[xshift=4.8cm, yshift=-3.8cm]{\sf malicious};

\node[ xshift=3.4cm, yshift=-3.3cm, rotate = 0,  text width=3cm, node distance=1cm, red]{\textcircled{\tiny 4} {\sf \footnotesize IP:4.3.2.1}};
\draw [-stealth,red,thick] (1.6,-3) -- (4,-3);

\node[bob,monitor, mirrored, minimum size=0.7cm, xshift=1.2cm, yshift=-3cm]{};
\node[xshift=1.2cm, yshift=-3.8cm]{\sf user};

\end{tikzpicture}
\caption{DNS cache poisoning attack. Attacker injects fake DNS entries into the DNS cache resolvers in an attempt to direct users into fake destinations, mainly for malicious purposes like phishing.}

\label{DNSCachePoisoning}
\end{center}
\end{figure}

The DNS resolver cache, or simply DNS cache, is a temporary database that stores resolved DNS look-ups. As a result, this caching enables users to quickly resolve a previously visited website. Unfortunately, DNS cache can be compromised through the insertion of unauthorized domain names and IP addresses. General flow of the DNS cache poisoning attack is illustrated in \autoref{DNSCachePoisoning}. In this manner, users' queries might be directed to a fake destination with malicious content or advertisements. DNS cache poisoning compromises the correct operation of Internet services and can be used for malicious activities, such as distributing malware and spam, phishing attacks, credential theft, \etc To this end, several research works have been conducted to detect and mitigate the DNS cache poisoning attacks~\cite{ShulmanW14, AntonakakisDLPLB10, PerdisciALL09, WuDZW15, ChenMP15, ChenAPNDL14, HaoW17}. 

\begin{table}[t]
\begin{center}
\small
\caption{Summary of the some of representative research works that have investigated DNS cache poisoning attacks. Here LDR is Low Deployment Rate of DNSSEC, AR is Acuracy Rate, and WSR donates }
\label{tab:SumCachePoisoning}
\scalebox{0.85}{
\begin{tabular}{l|l|l|c|c|c}
\Xhline{2\arrayrulewidth}

\multirow{2}{*}{\textbf{Work}} & \multirow{2}{*} {\textbf{Method}} & \multirow{2}{*}{\textbf{Advantages} } & \multicolumn{3}{c}{\textbf{Disadvantages}} \\ \cline{4-6}
                            &  & & LDR & AR & WSR \\
                            

\Xhline{2\arrayrulewidth}
\cite{ShulmanW14}          & DNSSEC                 & A-posteriori detection            & \xmark & &      \\ \hline
\cite{AntonakakisDLPLB10}    & Anax                   & Easy deployment                   & & \xmark &  \\ \hline
\cite{PerdisciALL09}            & WSEC DNS               & Easy deployment                   & & &  \xmark \\ \hline
\cite{HerzbergS12}        & DNSSEC                 & Cryptographic/verifiable          & \xmark & &     \\ \hline

\Xhline{2\arrayrulewidth}
\end{tabular}}
\end{center}
\end{table}

Herzberg and Shulman~\cite{HerzbergS12} studied the security of the patched DNS and found that not only source ports may be circumvented by various NAT devices, but also IP address randomization of standard-conforming resolvers can be circumvented. In addition, they demonstrated that DNS query randomization with both random prefix and 0x20 encoding can be circumvented easily. Finally, they suggested the deployment of DNSSEC as a countermeasure to prevent DNS cache poisoning attacks. Furthermore, Weaver \etal~\cite{WeaverKNP11} have analyzed a large number of measurement sessions from distinct IP addresses, collected by Netalyzr~\cite{KreibichWNP10}, to understand the DNS behavior. Their analysis revealed that DNS infrastructure suffers from significant limitations, such as inefficient look-ups, unreliability of IP-level fragmentation, and ISP-driven manipulation of DNS.

Shulman and Waidner~\cite{ShulmanW14} studied the security of DNS infrastructure and highlighted that although adoption of challenge-response defenses~\cite{RFC6056, RFC5452} are prevalent, DNS infrastructure is highly exposed to cache poisoning attacks. The authors claimed that DNSSEC is a suitable solution to defend against cache poisoning attacks.  
Antonakakis \etal~\cite{AntonakakisDLPLB10} have studied open recursive DNS resolvers' cache poisoning attacks and have found that attackers generally point victims to rogue IP addresses. Therefore, they have proposed Anax, a system that examines the nature of cache poisoning attacks and automatically detects them. Anax analyses resource records and extracts a set of statistical features, such as domain name diversity, 2LD diversity, 3LD diversity\footnote{xLD refers to the x-th level domain. In http://www.example.com, example.com is a 2nd level domain, and www.example.com is a 3rd level domain.}, \etc Then, these features are fed into a set of learning algorithms, such as SVM, neural network, \etc for cache poisoning attack detection. 

Perdisci \etal~\cite{PerdisciALL09} proposed Wildcard SECure (WSEC) DNS as a solution that defends against DNS cache poisoning attacks. WSEC DNS incorporates wildcard domains with TXT resource records to secure recursive DNS servers from cache poisoning attacks. WSEC DNS can be easily deployed, since it does not require any modifications of the root and TLD name servers. 

Wu \etal~\cite{WuDZW15} have proposed a DNS cache poisoning attack detection system based on the Kalman filter technique. The proposed system is composed of two subsystems: an entropy sequence, which is modeled using state space equations, and a second Kalman filter method, which is used to detect the attacks. The effectiveness of the proposed system is illustrated for both single and distributed cache poisoning attacks. The measurement errors and the correlation variation of the prediction errors are utilized for detection of the single and the distributed cache poisoning attacks, respectively. 


Chen \etal~\cite{ChenAPNDL14} have investigated the impact of disposable domains on the caching behavior of the DNS resolvers. Their analysis showed that as the prevalence of the disposable domains increase, it is likely that the DNS cache begins to be filled with resource records that are unlikely to be reused. In addition, their analysis revealed that disposable domains not only are widely used by different industries, such as Anti-Virus companies, \eg McAfee, Sophos, and popular websites, \eg Microsoft, Google, social networks, \eg Facebook, \etc, but also are increasing in that trend. 

Similarly, Hao and Wang~\cite{HaoW17} have studied the negative impact of one-time-use domain names on the performance of DNS caching. Therefore, they proposed a one-time-use domain detection system that incorporates syntactical features extracted from domain name string, such as length of query name, length of the longest sub-domain name, Sub-domain depth, \etc with machine learning algorithms is proposed. The authors argue that removal or even not inserting such resource records into the cache can prevent from waste of the DNS cache resources. 

\BfPara{Discussion} Different research works have proposed various methods for detection and mitigation of DNS cache poisoning attacks. These methods and their strengths and weaknesses are summarized \autoref{tab:SumCachePoisoning}. The limitations of proposed solutions can be summarized as:. 
\begin{enumerate*}
    \item low deployment rate of DNSSEC~\cite{HerzbergS12,ShulmanW14}, which by design address cache poisoning, calling for further exploration of how to creates incentives for the spread of DNSSEC deployment. 
    \item low detection accuracy~\cite{AntonakakisDLPLB10,HaoW17}, calling for better accuracy using multiple types of features. \item causing overheads on DNS traffic, memory usage and increasing response latency~\cite{PerdisciALL09}, calling for lightweight and more efficient approaches. 
    \item and wasting DNS cache storage resources by prefetching unpopular resource records~\cite{ChenMP15}, which calls for adaptive resolution. 
\end{enumerate*}

\subsection{Botnet and Attacks Using DNS}\label{sec:Botnet}

\begin{table}[t]
\begin{center}
\small
\caption{Summary of the research works that have investigated detection of botnet domains. Here AR is Accuracy Rate, while LDR donates Low Deployment Rate of DNSSEC.}
\label{tab:SumBotnet}
\scalebox{0.90}{
\begin{tabular}{l|l|l|c|c}
\Xhline{2\arrayrulewidth}
\multirow{2}{*}{\textbf{Work}} & \multirow{2}{*} {\textbf{Method}} & \multirow{2}{*}{\textbf{Advantages} } & \multicolumn{2}{c}{\textbf{Disadvantages}} \\ \cline{4-5}
                            &  & & AR & LDR \\

\Xhline{2\arrayrulewidth}
\cite{TruongC16}               & ML/domain length         & Small feature space               &\xmark & \\ \hline
\cite{JiangLLLDW12}                 & DNSSEC                                                                    & Strict delegation         & & \xmark \\ \hline
\cite{JiangLLLDW12}                   & ML/statistical features                              & Simple                             &\xmark & \\ \hline
\cite{XuBSY13}                         & Probability/DNS traffic                     & Simple                             &\xmark & \\
\Xhline{2\arrayrulewidth}
\end{tabular}}
\end{center}
\end{table}

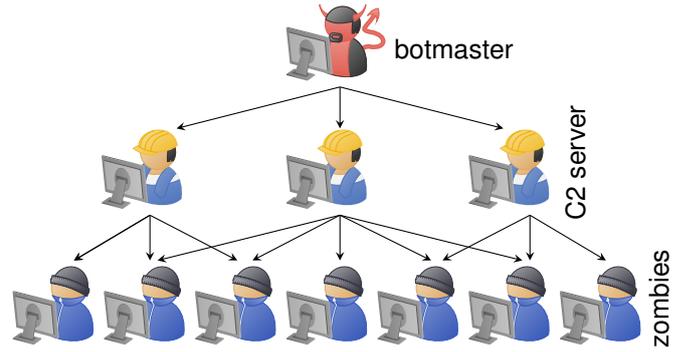
\begin{figure}[t]
\begin{center}
\begin{tikzpicture}

\node[devil,monitor, mirrored, minimum size=0.7cm, xshift=0cm, yshift=0.5cm]{};
\node[xshift=1.5cm, yshift=0.4cm]{\sf botmaster};

\node[builder, monitor, mirrored, minimum size=0.7cm, xshift=0cm, yshift=-1.2cm]{};
\node[builder,monitor, mirrored, minimum size=0.7cm, xshift=2.4cm, yshift=-1.2cm]{};
\node[builder,monitor, mirrored, minimum size=0.7cm, xshift=-2.43cm, yshift=-1.2cm]{};
\node[xshift=3.1cm, yshift=-1.1cm, rotate=90]{\sf C2 server};

\node[criminal,monitor, mirrored, minimum size=0.7cm, xshift=0cm, yshift=-3cm]{};
\node[criminal,monitor, mirrored, minimum size=0.7cm, xshift=1.2cm, yshift=-3cm]{};
\node[criminal,monitor, mirrored, minimum size=0.7cm, xshift=-1.2cm, yshift=-3cm]{};
\node[criminal,monitor, mirrored, minimum size=0.7cm, xshift=2.4cm, yshift=-3cm]{};
\node[criminal,monitor, mirrored, minimum size=0.7cm, xshift=-2.4cm, yshift=-3cm]{};
\node[criminal,monitor, mirrored, minimum size=0.7cm, xshift=3.6cm, yshift=-3cm]{};
\node[criminal,monitor, mirrored, minimum size=0.7cm, xshift=-3.6cm, yshift=-3cm]{};
\node[xshift=4.2cm, yshift=-2.9cm, rotate=90]{\sf zombies};

\draw [-stealth,black](0,-0.1) -- (0,-0.65);
\draw [-stealth,black](0,-0.1) -- (2.15,-0.65);
\draw [-stealth,black](0,-0.1) -- (-2.15,-0.65);

\draw [-stealth,black](0,-1.8) -- (0,-2.4);
\draw [-stealth,black](2.5,-1.8) -- (2.5,-2.4);
\draw [-stealth,black](0,-1.8) -- (1.2,-2.4);
\draw [-stealth,black](0,-1.8) -- (-1.15,-2.4);
\draw [-stealth,black](-2.5,-1.8) -- (-1.35,-2.4);
\draw [-stealth,black](-2.5,-1.8) -- (-2.5,-2.4);
\draw [-stealth,black](-2.5,-1.8) -- (-3.5,-2.4);
\draw [-stealth,black](-2.5,-1.8) -- (-3.5,-2.4);
\draw [-stealth,black](2.5,-1.8) -- (3.5,-2.4);
\draw [-stealth,black](0,-1.8) -- (-2.4,-2.4);
\draw [-stealth,black](0,-1.8) -- (2.4,-2.4);
\draw [-stealth,black](2.5,-1.8) -- (1.35,-2.4);

\end{tikzpicture}
\caption{An overview of a botnet infrastructure. Attacker (botmaster) remotely abuses infected devices (zombies) and its connections through C2 channels for malicious activities.}

\label{botnet}
\end{center}
\end{figure}

The word botnet is a combination of the words robot and network, and refers into a number of Internet-connected devices, each of which is running one or more {\em bots}. Botnets are composed of various infected hosts, C\&C channels, and a botmaster. The general infrastructure of a botnet is shown in \autoref{botnet}. Botnets can be used to perform DDoS attack~\cite{rustock}, password theft~\cite{riccardi2013titans}, and allow the attacker (botmaster) to access devices and their connection through C\&C channels. Researchers have investigated the threat of botnets various works, including the intersection between botnets and DNS~\cite{jinYGY2015, JiangLLLDW12, TruongC16, PlohmannYKBG16, XuBSY13}.

Jiang \etal~\cite{JiangLLLDW12} have studied whether deleting malicious domain names from upper level DNS servers can prevent botnet C\&C and malware propagation. Their analysis showed that ghost domain names stay resolvable even long after the delegated data has been removed from the domain registry and the record's TTL is expired. Finally, they suggest adoption of DNSSEC to overcome the threat of ghost domain attack. 

Jin \etal~\cite{jinYGY2015} have investigated the characteristics of the DNS log of botnet domain resolution and extracted six different features, \eg number of source IPs, total querying per day, and querying per hour, among others. These features are then fed into three different classification algorithms, including Adaboost, DT, and NB to detect botnet domain names. In addition, Chang \etal~\cite{ChangMWC17} have performed longitudinal study to identify novel botmaster strategies through analyzing multiple active botnet families. 

Truong and Cheng~\cite{TruongC16} have studied DNS traffic to design a detector that distinguishes domain names generated by legitimate users and pseudo-random domain names generated by botnets, such as Conficker~\cite{ThomasM14} and Zeus~\cite{MohaisenA13}. The proposed detector classifies the domain names based on machine learning algorithms, \eg RF, KNN, SVM, and NB, and using extracted features from DNS traffic, including length of domain names and their expected values. 

In addition, Xu \etal~\cite{XuBSY13} have carried out a comprehensive study to investigate the feasibility of DNS as a stealthy botnet C\&C channel.  Therefore, they have tested various strategies, \eg piggybacking query strategy and exponentially distributed query strategy, that can be utilized at the network level to effectively hide malicious DNS activities. Finally, a countermeasure is proposed that compares the probability distributions of legitimate DNS traffic and tunneling traffic. 

\BfPara{Discussion} Although various research works have studied botnets and possible solutions to defend against botnet attacks, such attacks are still compromising users and systems security, highlighting the limitations of the proposed methods. These methods and their strengths and weaknesses are summarized in  \autoref{tab:SumBotnet}. The shortcomings of the proposed methods can be summarized as follows: Low detection accuracy~\cite{TruongC16, JiangLLLDW12}, high false alarm rate~\cite{XuBSY13}, and low deployment rate of DNSSEC~\cite{JiangLLLDW12}. In addition, the emergence and rise of IoT and associated vulnerabilities opens a new research direction that requires further investigations. For example, the Mirai botnet attacks insecure IoT devices, \eg routers and digital video recorders~\cite{whittaker_2016, palmer_2018} highlighting the need for security solutions optimized for the IoT devices~\cite{kolias2017ddos, bertino2017botnets}

\subsection{Phishing Attacks and Defenses}\label{sec:Phishing}

Attackers in phishing attack, attempt to acquire  personal and secret information, such as passwords and details of credit cards, by disguising as a trustworthy entity in an electronic communication. In light of the MCSI review published on 2014, the yearly overall effect of the phishing and different types of data fraud attacks has been assessed at as high as US\$5 billion \cite{MicrosoftNews} only at one particular country, Singapore. Therefore, researchers have conducted a large number of works to detect and mitigate phishing attacks \cite{BilgeKKB11, DagonPLL08,KintisMLCGPNA17, CookGD08,MedvetKK08} .

For example, Bilge \etal~\cite{BilgeKKB11} have presented a malicious domain detection system, called EXPOSURE, which analyzes various characteristics of DNS names and the way that they are queried. EXPOSURE analyzes DNS traffic and extracts different types of features, including time-based feature, DNS answer-based features, TTL value-based features, and domain name-based feature. Those features are then used by EXPOSURE to design a decision based classifier for automatic detection of a wide range of malicious domain names, \eg botnet C\&C servers, phishing sites, and scam hosts. 

\begin{table}[t]
\begin{center}
\small
\caption{Summary of sample research works that have studied phishing attacks. Here G is Generic, L is Lightweight, and AR donates Accuracy Rate. . The common weakness of these studies is low detection accuracy. }
\label{tab:SumPhishing}
\begin{tabular}{l|l|c|c|l}
\Xhline{2\arrayrulewidth}
\multirow{2}{*}{\textbf{Work}} & \multirow{2}{*} {\textbf{Method}} & \multicolumn{2}{c|}{\textbf{Advantages}} & \multirow{2}{*}{\textbf{Disadvantages}} \\ \cline{3-4}
                            &  & G & L & \\
\Xhline{2\arrayrulewidth}
\cite{BilgeKKB11}                 & EXPOSURE                          & \cmark &             & AR \\ \hline
\cite{CookGD08}                    & Phishwish                         & & \cmark             & AR \\ \hline
\cite{MedvetKK08}                & Visual similarity                 & & \cmark             & AR \\\hline
\cite{HaraYM09}                    & Visual similarity                 & \cmark  &            & AR\\ 
\Xhline{2\arrayrulewidth}
\end{tabular}
\end{center}
\end{table}

\BfPara{Web-based approaches}
Cook \etal~\cite{CookGD08} proposed a mechanism, Phishwish, to detect phishing messages or emails with a small false positive rate. The idea of Phishwish was to provide better protection against zero-hour attacks than blacklists. Phishwish analyzes text and HTML formatted emails using four general rules: identification and analysis of the log-in URL in the email, analysis of the email headers, analysis of the images in the email, and analysis of the accessibility of the URLs.

Medvet \etal~\cite{MedvetKK08} presented a phishing detection method based on the visual-similarity that works based on analyzing three main characteristics of websites; the text pieces, the embedded images in the page, and the overall visual appearance of the website. A similarity signature is calculated for legitimate and suspicious websites and an alarm is raised if the signatures were too similar. 
Furthermore, Hara \etal \cite{HaraYM09} have proposed a phishing detection technique based on visual similarity that works even if the original website is not registered in the database. A collection of legitimate websites are used to train the classifier and store in a database. A suspected website's snapshot is then compared to the websites in the database; the suspected website is labeled as phishing if the image similarity metric is above a certain threshold. In case where there is no similar website in the database the suspected website is considered as legitimate website.  

\BfPara{Discussion} Despite the large number of research works that have studied the detection and mitigation of phishing attacks, the landscape of phishing is an expanding area contributing very much to the DNS security landscape. \autoref{tab:SumPhishing} summarizes the proposed methods in the literature and their strengths and weaknesses. Low detection accuracy~\cite{CookGD08,HaraYM09}, requiring large number of train samples~\cite{BilgeKKB11}, dependency of the performance to train inputs~\cite{BilgeKKB11, MedvetKK08}, and not being adaptive to the changes in the scenarios~\cite{HaraYM09} are some of the shortcomings of the proposed methods in the literature. In addition, phishers are targeting mobile messaging and social apps in increasing rate~\cite{wandera_2018}, which requires further detailed investigations to be addressed.

\subsection{DNS Manipulation}\label{sec:Manipulation}

The process of diverting legitimate DNS requests to malicious IP addresses, which are hosted on misbehaving servers, is known as DNS manipulation. DNS manipulation behavior of attackers exposes users to threats, such as phishing and content injection. It is thus not a surprise that a large number of studies have focused on DNS manipulation \cite{SchompCRA14, KuhrerHBRH15, PearceJLEFWP17, TrevisanDMM17}. 

For instance, the vulnerability of the user-side DNS infrastructure to record injection threats have been measured by Schomp \etal~\cite{SchompCRA14}. Their analysis showed that a large number (9\%) of open DNS resolvers are vulnerable to record injection attacks and are abused by attacks on shared DNS infrastructure. They have measured the extent of popular record injection attacks, \eg Kaminsky~\cite{kaminsky2008black} and the deployment of familiar defensive methods, \eg 0x20 encoding~\cite{DagonAVJL08}.


Kuhrer \etal~\cite{KuhrerHBRH15} have investigated the negative aspects of open DNS resolvers that can be abused by attackers for various malicious activities, \eg amplification DDoS attacks, DNS manipulation, and cache poisoning, among others. The authors have investigated the response authenticity of the open resolvers from the user's point of view, and found that millions of them deliberately manipulate DNS resolutions for censoring communication channels, injecting advertisements, serving malicious files, and performing phishing attacks.  Furthermore, Jones \etal~\cite{JonesFPWA16} have proposed an approach for the detection of unauthorized DNS root manipulation using two different techniques. The first technique analyzes the latency to root servers and the second technique analyzes route hijacks. Their analysis revealed that the entities that are operating unauthorized root servers can completely control the entire Internet name space for any system within their sphere. 

Pearce \etal~\cite{PearceJLEFWP17} proposed a scalable and lightweight system, called Iris, which measures and detects the widespread of DNS manipulation at the scale of countries, which were manipulated based on tactics that rely on DNS resolvers. Iris collects DNS queries through geographically distributed DNS resolvers and analyzes the responses based on two metrics, the consistency and the independent verifiability metrics. Iris investigates both sensitive domains and domains of popular websites for DNS manipulation. Their analysis showed that DNS manipulation is a phenomenon that is heterogeneous across resolvers, domain names, and countries.

Trevisan \etal~\cite{TrevisanDMM17} proposed an automatic and parameter-free detection system, called REMeDy, which identifies rogue DNS resolvers. REMeDy analyzes DNS traffic and evaluates the consistency of responses across all resolvers for automatic identification of manipulated responses.

\BfPara{Discussion} DNS infrastructure suffers from multiple issues, \eg inefficient look-ups, unreliability of IP-level fragmentation, and ISP-driven DNS manipulation~\cite{WeaverKNP11}. Moreover, record injection vulnerabilities, \eg Kaminsky~\cite{kaminsky2008black} are widespread \cite{SchompCRA14}. Attackers can utilize standard-conforming resolvers to circumvent mitigation techniques \eg source port randomization as well as IP address randomization~\cite{HerzbergS12, TrevisanDMM17} to manipulate DNS. In addition, ISPs are clearly willing to involve in DNS manipulation for reasons, such as error traffic monetization~\cite{WeaverKP11}.  Furthermore, it has been demonstrated that root servers can be abused by unauthorized entities for malicious activities, such as blocking access to websites or manipulate responses through MITM proxies~\cite{JonesFPWA16}.

\begin{figure}[t]
\begin{center}
\includegraphics[width=0.45\textwidth]{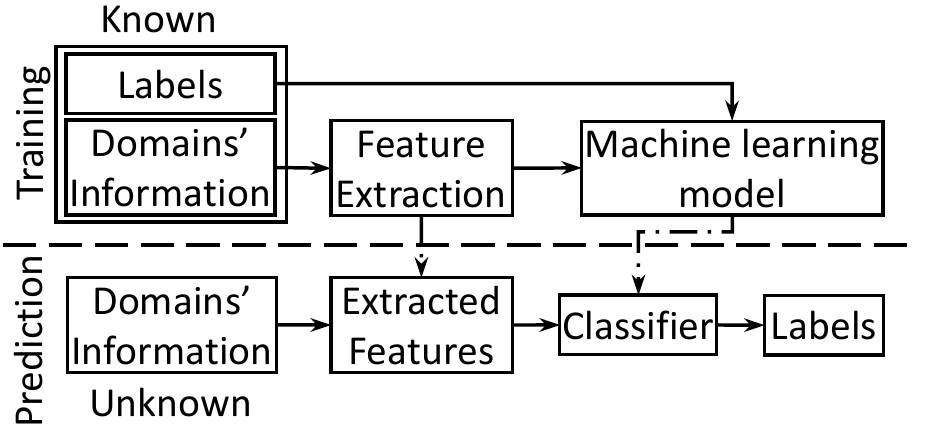}
\caption{General flow of the malicious domains detection based on machine learning algorithms. It is composed of two steps; training ML model over representative features of labeled domain names and then using the trained model to classify unknown domain names.}
\label{MDD_ML}
\end{center}
\end{figure}

\begin{table}[t]
\begin{center}
\caption{Summary of the research works that have investigated malicious domain detection. Here G is Generic, L is Lightweight, P is Proactive, AR is Accuracy Rate, E is Evadable, DID is Dependency on Input Data, and LV donates Local Visibility. }
\label{tab:SumMaliciousDomains}
\scalebox{0.850}{
\begin{tabular}{l|l|c|c|c|c|c|c|c}
\Xhline{2\arrayrulewidth}
\multirow{2}{*}{\textbf{Work}} & \multirow{2}{*} {\textbf{Method}} & \multicolumn{3}{c|}{\textbf{Adv.}} & \multicolumn{4}{c}{\textbf{Disadv.}}  \\ \cline{3-9}
                            &  & G & L & P & AR & E & DID & LV\\
\Xhline{2\arrayrulewidth}
\cite{jinYGY2015}                  & ML/statistical       & &\cmark  &              & \xmark & & &       \\ \hline

\cite{BilgeKKB11}                & EXPOSURE                      & \cmark & &              & & \xmark &\xmark & \xmark                 \\ \hline

\cite{AntonakakisPDLF10}   & Notos                         & & & \cmark              & \xmark & & \xmark & \xmark                  \\ \hline
\cite{AntonakakisPLVD11}   & Kopis                         & & \cmark &              & & & \xmark &       \\ \hline
\cite{BilgeSBKK14}               & Ext. EXPOSURE             & & \cmark &              & & \xmark &\xmark & \xmark   \\ \hline
\cite{CanaliCVK11}              & Prophiler                     & &\cmark  &              & \xmark & & &        \\ \hline
\cite{FelegyhaziKP10}       & Proactive blacklisting        & & &\cmark               & \xmark & & &              \\ \hline
\cite{GaoYCPGJD13}                 & Temporal correlation          &\cmark &  &              & & \xmark & &     \\ \hline
\cite{HaoKMPF16}                  & PREDATOR                      & & & \cmark              & \xmark & \xmark & &      \\ \hline
\cite{LuoTZSLNM15}                 & Syntactic/temporal   & \cmark & &              &  & \xmark & &       \\ \hline
\cite{LiuLDWLD17}                 & Woodpecker                    & & & \cmark              & & \xmark & &   \\ 
\Xhline{2\arrayrulewidth}
\end{tabular}}
\end{center}
\end{table}

\subsection{Malicious Domains Detection}\label{sec:MaliciousDomain}

Domains and domain names provide a hierarchy of unique identifiers that guide traffic across the Web and identify websites, servers and other resources. Notwithstanding, in the form of malicious domains, they are an essential apparatus in the hands of miscreants. There are no silver bullets for securing against malicious domains. Nonetheless, understanding domain names can enable firms and individual stakeholders to protect themselves against attacks. A wide variety of research works have been done for accurate detection of malicious domain names\cite{AntonakakisPLVD11, BilgeKKB11, BilgeSBKK14, CanaliCVK11, FelegyhaziKP10, GaoYCPGJD13, HaoKMPF16, HaoTPFKGH13, jinYGY2015, LuoTZSLNM15, MaSSV09,HaoFP11,LiuLDWLD17}. 

This work is classified into two general categories, classification based~\cite{AntonakakisPDLF10,AntonakakisPLVD11,BilgeKKB11,BilgeSBKK14,jinYGY2015} and inference based~\cite{GaoYCPGJD13,LuoTZSLNM15,LiuLDWLD17}. The classification based approaches rely on local network and host information; however, inference based approach exploits the global information along with local information. The general structure of the malicious domain detection systems based on classification approaches is shown in \autoref{MDD_ML}.  

\BfPara{Classification based approach} Antonakakis \etal~\cite{AntonakakisPDLF10} have presented a malicious domain detection system, Notos, which works based on assigning dynamic reputation score to any domain name in a DNS query. Notos maintains up-to-date DNS information of domain names by gathering such information from various sources such as the DNS zone and associated IP addresses, BGP prefixes, AS information, and honeypot analysis. This information is then used to build models of benign/malicious domain names, and then used to calculate the reputation scores of new domains indicative of their maliciousness. 

Instead of monitoring traffic from local recursives, Antonakakis \etal~\cite{AntonakakisPLVD11} presented Kopis, which detects malware-related domains by monitoring DNS traffic at the upper DNS hierarchy. Kopis offers a new set of traffic features, including daily patterns of domain name resolution, the significance of requesters within each epoch, and the reputation of IP addresses that leverage a global visibility, thus leading to an early detection of malicious domains. Kopis offers independent malware domain detection within DNS operator's authority and works well even if there is no available information about IP reputation.

Bilge \etal~\cite{BilgeKKB11} have presented a malicious domain detection system, called EXPOSURE, which analyzes a large set of passive DNS records. EXPOSURE analyzes various characteristics of DNS names and the way that they are queried, and extracts various kinds of features, namely time-based feature, DNS answer-based features, TTL value-based features, and domain name-based feature. These features are then used to design a decision tree based classifier that automatically detects a wide range of malicious domain names, \eg botnet C\&C servers, phishing sites, and scam hosts. Later they extended their work by analyzing a large set of real-world DNS requests over 17 months period of EXPOSURE's operation~\cite{BilgeSBKK14}. They showed that EXPOSURE is able to make meaningful connections between various malicious domains. For example, there exist IP addresses that are shared among thousands of unique domain names indicative of botnet-related domain names. 

Jin \etal~\cite{jinYGY2015} have investigated the characteristics of the DNS log of botnet domain resolution to extract representative features, \eg Number of source IPs, Total querying per day, Querying per hour \etc These features are then used to build a malicious domain detection system that consists of three different classification algorithms: Adaboost, DT, and NB.  


\BfPara{Inference-based approach} Strong associations of domains with known malicious domains is a vital point in effective malicious domain detection. Therefore, Khalil \etal~\cite{KhalilGNY18} have designed an association-based scheme for the detection of malicious domains with high accuracy and coverage. They have analyzed active DNS data and extracted two types of features, namely, domain based features and IP block based features. These features are then used to accurately distinguish public IPs from dedicated ones, consequently building high-quality associations between domain names and identifying malicious domains.  
Furthermore, Gao \etal~\cite{GaoYCPGJD13} have utilized temporal correlation analysis of DNS queries to detect unknown malicious domains based on the related known malicious anchor domains. The proposed approach is able to detect a wide range of correlated malicious domain groups, \eg phishing, spam, and DGA-generated domains. The results demonstrated that on average each of the known malicious anchor domains can detect more than 53 previously unknown malicious domains.

Luo \etal~\cite{LuoTZSLNM15} have proposed a comprehensive framework that uses evolutionary learning for detection of diverse clusters of DNS failures, \eg highly random, mutated string, and sub-string domain name failure patterns using different syntactic and temporal patterns. In order to reduce the computational cost, less suspicious clusters were removed while more suspicious cases were preserved for further analysis. 

\BfPara{Time-of-registration detection} Early detection of potentially malicious DNS domains are of high importance for network operators, registries or registrars, and law enforcement professionals to defend against many Internet attacks. Therefore, Hao \etal~\cite{HaoTPFKGH13} have explored registration behaviors of malicious domains with an eye towards features that might indicate the maliciousness of a domain at the time of registration. The characteristics of registrars, domains life cycle, and names patterns were investigated and the results showed that spammers use a small set of registrars to register previously used domains in bulk. 

Hao \etal ~\cite{HaoKMPF16} have proposed a malicious domain detection system, PREDATOR, that proactively detects malicious domains using time-of-registration statistical features, including registration history features, domain profile features, and batch correlation features. These features are used to build ML-based classifier, \eg SVM, CMP that detects malicious domains before they appear in DNS blacklists. Moreover, Felegyhazi \etal~\cite{FelegyhaziKP10} proposed an approach for proactive malicious domain blacklisting that relies on registration and name server information of small set of known malicious domains to predict whether a larger set of domains are malicious. Their analysis showed that on average 3.5 to 15 new domains can be derived from a given known malicious domain. 

\BfPara{Web-based approach} Ma \etal~\cite{MaSSV09} have studied application of batch and online learning approaches for detection of malicious web sites using live feed of labeled URLs and based on the lexical and host-based features. Their analysis showed that size of the training dataset and changes on the distribution of the features affects the performance of the batch algorithms. However, online confidence weighted algorithm is able to classify the labeled URLs in real time and with high accuracy.
Furthermore, Canali \etal~\cite{CanaliCVK11} have proposed a malicious web pages detection system based on a fast and reliable filter, Prophiler, that filters out benign web pages from further costly analysis. Prophiler examines a web page for malicious contents using different types of static features, such as HTML features, JavaScript features, URL and host-based features. 

\BfPara{Discussion} Although there exists a large body of research devoted to detection of malicious domain names, miscreants have used the DNS to build malicious network infrastructure, \eg botnet~\cite{PerdisciCG12}. The summary of the proposed methods and their strengths and weaknesses are listed in~\autoref{tab:SumMaliciousDomains}. Shortcomings of the proposed methods can be summarized as following: low detection accuracy~\cite{AntonakakisPDLF10,CanaliCVK11,jinYGY2015,FelegyhaziKP10,HaoKMPF16}, local visibility~\cite{AntonakakisPDLF10,BilgeKKB11,BilgeSBKK14}, expensive computation~\cite{MaSSV09}, dependency on training dataset~\cite{AntonakakisPDLF10,BilgeKKB11,BilgeSBKK14,AntonakakisPLVD11}, requiring correlation between anchor malicious domain and unknown malicious domains~\cite{LuoTZSLNM15,GaoYCPGJD13}, and can be evaded by avoiding certain features or behaviors~\cite{BilgeKKB11,BilgeSBKK14,HaoKMPF16}.

\begin{figure}[t]
\begin{center}
\includegraphics[width=0.5\textwidth]{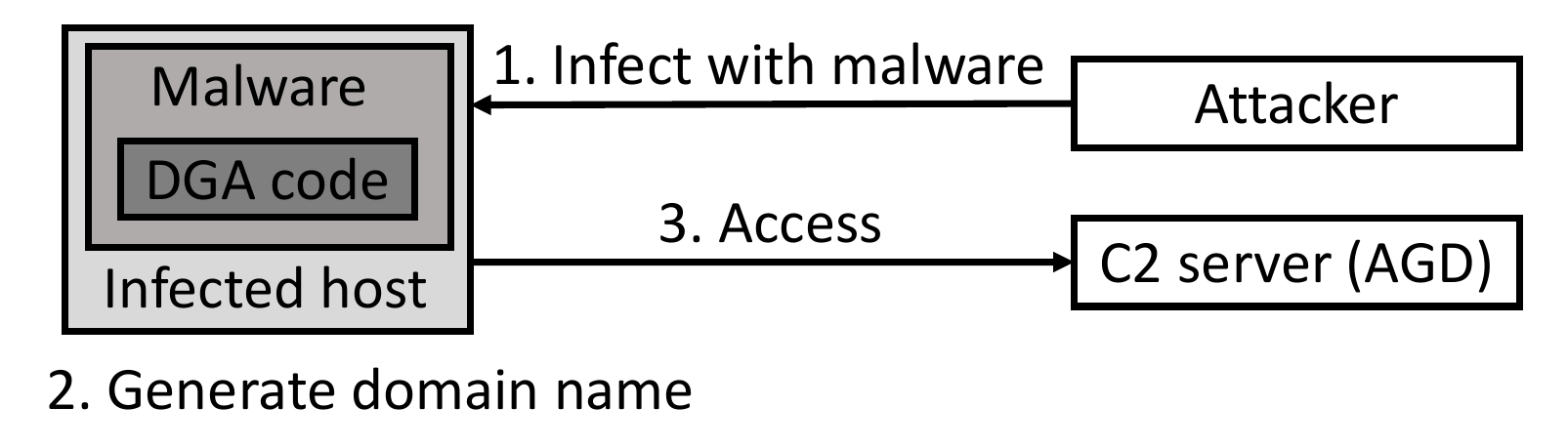}
\caption{The General overview of the communication of the malware-infected hosts with C\&C servers. DGAs are being used as main method of communication in botnets.}
\label{DGAStruct}
\end{center}
\end{figure}

\subsection{Domain Generation Algorithm}\label{sec:DGA}
Malware operation sometimes relies on a fixed domain or IP addresses, which can be hindered quickly. Thus, the primary idea of Domain generation algorithms (DGAs) were instead of developing new versions of the malware or setting everything up at another server, the malware would transfer its C\&C to another domain at regular intervals. DGAs are utilized to intermittently produce a large number of domain names and usually for malicious purposes. For instance, \autoref{Fig:DGA} shows a sample DGA code that is utilized to generate algorithmic domain names. DGAs are being used as the main method of communication in botnets. For example, \autoref{DGAStruct} shows an overview of the communication of malware-infected hosts with C\&C servers. There exist several research works that explored the ecosystem of DGAs~\cite{YadavRRR10, AntonakakisPNVALD12, baraboschWLG12, PlohmannYKBG16, TruongC16, SpauldingPKM18}. 

Yadav \etal~\cite{YadavRRR10} studied algorithmically generated domain names for domain fluxing. Statistical measures, such as Kullback-Leibler divergence~\cite{kullbackL1951}, Jaccard index, and the Levenshtein edit distance were used to explore whether a group of domains are algorithmically generated or not. They have applied their method to Tier-1 ISP's DNS traces and the results showed that it can automatically detect domain fluxing, even for unknown and unclassified botnets, \eg Mjuyh.

Plohmann \etal~\cite{PlohmannYKBG16} have conducted a large-scale measurement study to understand domain generating malware. They investigated and presented a taxonomy for 43 DGA-based malware families to identify and compare their characteristics. Using reverse engineering and re-implementation of the algorithms they pre-computed more than 159M possible DGA-based domain names, which can be used for both predictive blocking of C\&C accesses as well as accurate identification of malware families and related campaigns of future DGA domain names.

\begin{figure}[t]
{\tt
\begin{lstlisting}[language=Python]
def generate_domain(y, m, d):
   domain = “”
   for i in range(16):
       y=((y^8*y)>>11)^((y&0xFFFFFFF0)<<17)
       m=((m^4*m)>>25)^16*(m&0xFFFFFFF8)
       d=((d^(d<<13))>>19)^((d&0xFFFFFFFE)<<12)
       dm+=chr(((y^m^d)%25)+97)
   return domain
\end{lstlisting}
}
\caption{A sample DGA code that is utilized to intermittently generate large number of algorithmic domain names, mainly for malicious purposes~\cite{SpauldingPKM18}.}
\label{Fig:DGA}
\end{figure}

Antonakakis \etal~\cite{AntonakakisPNVALD12} have investigated unsuccessful DNS resolutions to detect groups of potential DGA domains. They built a DNS-based DGA domain detection system, called Pleiades, which works based on the characteristics of partially registered algorithmically generated domain names. 

Spaulding \etal~\cite{SpauldingPKM18} have presented a malicious domain detection system that proactively detects algorithmically generated domain names. Their work  highlighted the difference between the number of NXDomain responses of benign and algorithmically generated domain names prior to registration. These patterns are then used to build a classification system using sliding time windows. Barabosch \etal~\cite{baraboschWLG12} have presented an automatic method to extract DGAs from malware binaries using dynamic analysis in combination with data flow analysis. In addition, a taxonomy of DGA families is defined using time dependency and causality features.  
Truong \etal~\cite{TruongC16} have analyzed DNS traffic to design a detection system for recognition of algorithmically generated domain names form legitimate domain names. The length of domain names and their expected values are then used to build a classifier that identifies pseudo-random domain names generated by botnets, such as Conficker~\cite{ThomasM14} and Zeus~\cite{MohaisenA13}.

\BfPara{Discussion} Some major challenges of detecting DGA-based domains include constructing a taxonomy of DGA families~\cite{baraboschWLG12,PlohmannYKBG16}, collecting domain names generated by different malware families~\cite{PlohmannYKBG16}, reconstruction of DGAs by reverse engineering malware\cite{AntonakakisPNVALD12,baraboschWLG12}, learning newly found DGAs (being generic)~\cite{AntonakakisPNVALD12,YadavRRR10}, and low detection accuracy~\cite{TruongC16}.  

\subsection{Domain Name Squatting} \label{sec:Squatting}
Domain name squatting, also known as cybersquatting, is registering or using an Internet domain name with the bad intent to profit from a trademark belonging to someone else. Cybersquatters register variants of popular trademark names through different squatting strategies, including typosquatting, bitsquatting, combosquatting, and soundsquatting. Domain name squatting practices expose users to a variety of vulnerabilities, such as trademark infringement, monetization, malware, and scams. There exists several research works that attempted to investigate the landscape and impact of the domain squatting~\cite{AgtenJPN15, KhanHLK15, KintisMLCGPNA17, SpauldingNM17, SzurdiKCSFK14, SpauldingUM17, VissersBGJN17}, which we define below.

\subsubsection{Typosquatting} 
\begin{figure}[t]
\begin{center}
\includegraphics[width=0.5\textwidth]{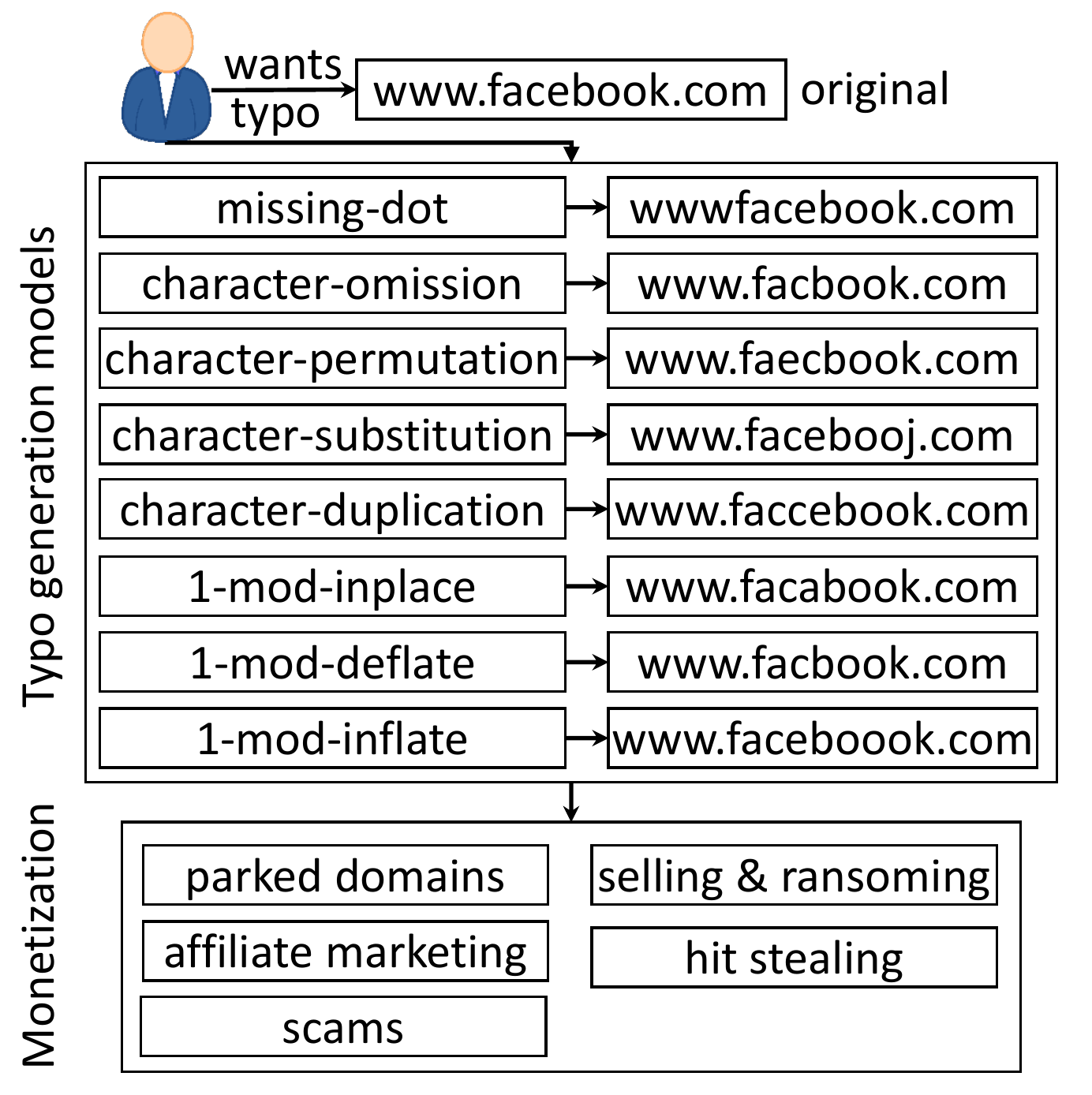}
\caption{Common typo-generation models and monetization strategies. Notice that permutations and substitutions of characters are among the most effective techniques in deceiving users.}
\label{fig:Typosquatting}
\end{center}
\end{figure}

Typosquatting is one of the most common forms of domain squatting that targets Internet users who incorrectly type a website address into their web browser, \eg \textit{www.examlpe.com} instead of \textit{www.example.com}, and mainly for monetization activities. To this end, researchers have attempted to systematically introduce typo-generation models~\cite{DWangBWVD06,BanerjeeBFB08}. For instance, Wang \etal~\cite{DWangBWVD06} have discovered common typo-generation models of Alexa top domain list, namely missing-dot typos, character-omission typos, character-permutation typos, character-substitution typos, and character-duplication typos. Similarly, Banerjee \etal~\cite{BanerjeeBFB08} have suggested different methods for generating typosquatting domains, including 6.1-mod-inplace, 7.1-mod-deflate, and 8.1-mod-inflate. Furthermore, Spaulding \etal~\cite{SpauldingUM16} have explored the landscape of the typosquatted domain names and summarized common typo-generation models and  monetization strategies, which are shown in~\autoref{fig:Typosquatting}.

Agten \etal~\cite{AgtenJPN15} conducted a longitudinal study of typosquatting domains and trends. Their analysis showed that 95\% of the studied top domains are targeted by typosquatters; however, only few trademark owners take protection practices, such as proactive registration of typosquatted domains. Furthermore, they observed a change in the trend and monetization behaviors of typosquatters over time and through hosting different types of websites, \eg parked/ads/for sale domains and scams. 
Moreover, Szurdi \etal~\cite{SzurdiKCSFK14} conducted a large-scale measurement of typosquatting domain registrations in the .com TLD. This study explores the monetization strategies among less known domains. Their analysis showed that 95\% of typosquatting domains target the long tail of the popularity distribution. In addition, they found that typosquatting domains constitute 20\% of the domains in the .com TLD and the trend is increasing. Moreover, \etal~\cite{BanerjeeBFB08} revealed that for a given authoritative domain the percentage of active typosquatting domains decreases remarkably as the popularity of domains decline. 

However, the results of these two research works~\cite{SzurdiKCSFK14, AgtenJPN15} are in contrast to the results presented by 

Khan \etal~\cite{KhanHLK15} have explored the impact of typosquatted domains on the users through measuring the harm experienced by users using intent inference metric. Their analysis showed that typosquatting practice increases the time it takes to find an intended website, generally 1.3 seconds per typosquatting event over the alternative of receiving a browser error page. 

Spaulding \etal~\cite{SpauldingNM17, SpauldingUM17} have conducted a user study to understand the effectiveness of the typosquatting strategies in deceiving users. The behaviour of users, who have been exposed to several typosquatting URLs, have been examined in an attempt to figure out whether their behavior is improved by security education and increasing their awareness about typosquatters. They found that user's behavior is affected the most by factors such as age and education. Furthermore, the results showed that typo-generation models, such as permutations and substitutions of characters are among the most effective techniques in deceiving users, as opposed to techniques that rely on adding characters to the domain names.

\subsubsection{Bitsquatting} The main premise of bitsquatting~\cite{NikiforakisAMDPJ13,Dinaburg} is hardware malfunction and it refers to registration of domain names with one different bit from popular domain names in an attempt to receive unintentional traffic caused by bit-flip errors in the memory of computers. This theory is tested, for the first time, by Dinaburg~\cite{Dinaburg} through investigating all HTTP requests to 30 registered bitsquatted versions of popular domain names, \eg amazon.com. The results of  eight months of experience showed that there were over 52K bitsquat requests from more than 12K unique IP addresses, highlighting the feasibility and presence of bitsquatting. Finally, a hardware-based defense strategy that analyzes the integrity of data stored in hardware is introduced. Furthermore, Nikiforakis \etal~\cite{NikiforakisAMDPJ13} explored the bitsquatting phenomenon in more detail and found that attackers register new bitsquatting domains daily. Their results showed that bitsquatting domains involve different monetization practices, such as parked/ads/for sale domains, affiliate marketing, and scams. Finally, they reviewed possible defense strategies against bitsquatting, such as data integrity validation, DNSSEC, and pre-registering all possible bitsquatting domains. 

Vissers \etal~\cite{VissersBGJN17} have conducted a large-scale analysis of configuration issues, \eg typosquatting and hardware errors, \eg bitsquatting over 10K popular name server domains. Their analysis confirmed the presence of squatting vulnerabilities among name servers, thus compromising all domains that rely on them. Furthermore, they analyzed the security practices of popular name servers and found that 12K domains are directly exposed to being hijacked through configuration errors related to their name server and 52.8M domains are being targeted by name server bitsquatters. Finally, they suggested deployment of DNSSEC as an effective countermeasure. 


\begin{table*}[t]
\begin{center}
\caption{Summary of the works that have studied the landscape of the domain name squatting. T= Typosquatting, B= Bitsquatting, C= Combosquatting, SS= Soundsquatting, PAF= Parked/Ads/For sale domains, S= Scams, HS= Hit Stealing, and AM= Affiliate Marketing. }
\label{tab:SumSquatting}
\begin{tabular}{l|c|c|c|c|c|c|c|c|c|c}
\Xhline{2\arrayrulewidth}
\multirow{2}{*}{\textbf{Work}} & \multicolumn{4}{c|}{\textbf{Method}} & \multicolumn{4}{c|}{\textbf{Monetization strategy}} & \multirow{2}{*}{\textbf{Contribution}} & \multirow{2}{*}{\textbf{Disadvantages}}  \\ \cline{2-9}
                            &  T & B & C & SS & PAF & S & HS & AM &  & \\                
\Xhline{2\arrayrulewidth}
\cite{SpauldingUM16}    & \cmark  & \cmark &  & \cmark            & \cmark & \cmark & \cmark & \cmark  & Squatting comparison & --- \\ 
\hline
\cite{KintisMLCGPNA17}    &  &  & \cmark &         &  &   & & \cmark  & Combosquatting & No detection  \\ \hline

\cite{DWangBWVD06}        & \cmark & & &          & \cmark & & &  & Typo-generation models & Limited data \\ \hline

\cite{AgtenJPN15}        & \cmark & & &          & \cmark & \cmark & &  & Longitudinal study & Limited data   \\ \hline

\cite{SzurdiKCSFK14}    & \cmark & & &      & \cmark & & & \cmark & YATT mitigation tool  & Updating blacklist   \\ \hline

\cite{KhanHLK15}          & \cmark & & &      & \cmark & & &   & Typosquatting detector  & Large user traffic   \\ \hline

\cite{Dinaburg}       & & \cmark & &         & \cmark & \cmark & & \cmark   & Bitsquatting/defense & impractical \\ \hline

\cite{NikiforakisAMDPJ13}    & & \cmark & &           & \cmark & \cmark & & \cmark  & Bitsquatting/defense & Costly/impractical\\ \hline

\cite{VissersBGJN17}    & \cmark & \cmark & &    &  & \cmark  & &  & Name server squatting & Low deployment\\ \hline

\cite{NikiforakisBDPJ14}    & &  & & \cmark         & \cmark & & \cmark & \cmark & Soundsquatting & Accuracy rate \\ 
\Xhline{2\arrayrulewidth}
\end{tabular}
\end{center}
\end{table*}

\subsubsection{Combosquatting} Combosquatting is a type of domain squatting in which a popular trademark is combined with one or multiple phrases, \eg \textit{www.youtube-login.com} instead of \textit{www.youtube.com}. In order to explore its scope, Kintis \etal~\cite{KintisMLCGPNA17} have analyzed Lexical characteristics and temporal properties of DNS records, collected from PDNS and ADNS sources over six years, for in depth and long-term analysis of combosquatting domains. The results showed that combosquatting domains are generated by adding a single token to the original domain and they are far more prevalent than typosquatting. In addition, they found that combosquatting domains are increasing in numbers, while large fraction of them ($\approx$60\%) stay active for more than 1000 days.

\subsubsection{Soundsquatting} Soundsquatting is a type of domain squatting technique that takes the advantage of homophones words and the users confusion, such as such as \textit{{whether,
weather}} and \textit{{idle, idol, idyll}} to generate squatted domains.  Nikiforakis\etal~\cite{NikiforakisBDPJ14} have explored the landscape of soundsquatting and showed that attackers are already familiar with soundsquatting concepts, \eg generative models of soundsquatted domains and are monetizing them in different ways, such as parked/ads/for sale domains, hit stealing, and affiliate marketing. Furthermore, they proposed a tool, Auto Sound Squatter (AutoSS), that automatically generates soundsquatting domains. In addition, they showed that users are exposed to soundsquatting through the abuse of text-to-speech software, including the built-in screen reader of Windows XP, Windows 7 and Mac OS X, the Thunder screen reader~\cite{Thunder}, ORCA~\cite{Orca}, and Skyvi~\cite{skyvi}. 

\BfPara{Discussion} Researchers have identified various domain squatting strategies \eg  typosquatting, bitsquatting, \etc, which are listed in~\autoref{tab:SumSquatting} along with the summary, contribution, and limitations of the research works that investigated them. Despite the proposed methods to mitigate domain squatting, squatting phenomenon is continuing to thrive and expand~\cite{SzurdiKCSFK14,KintisMLCGPNA17}. The limitations of the existing research proposals can be summarized as following: 1) lack of practical yet automatic domain squatting detection system~\cite{KintisMLCGPNA17}, 2) requiring a huge amount of user traffic or content for building models~\cite{KhanHLK15}, and 3) lack of practical yet effective mitigation techniques~\cite{NikiforakisAMDPJ13, VissersBGJN17, Dinaburg}. For instance, the research proposals which suggest deployment of DNSSEC~\cite{NikiforakisAMDPJ13, VissersBGJN17} should consider the fact that deployment rate of DNSSEC although increasing but still far from ideal point. 

\subsection{Parked Domain Monetization}\label{sec:Monetization}

Domain parking refers to the registration of an Internet domain name without having the domain  associated with any services such as e-mail or a website. Domain parking can be classified as monetized and non-monetized. In the former, advertisements are shown to visitors and the domain is monetized; whereas, non-monetized parked domains reserve the domain name for future development or to protect against the possibility of cybersquatting. Despite technically being a legitimate business, parked domain monetization can get mixed up with suspicious practices and malicious contents such as malware. Therefore, several works were conducted to understand  parked domains and monetization strategies \cite{AlrwaisYALW14, SzurdiKCSFK14, WeaverKP11, VissersJN15}. 

Alrwais \etal~\cite{AlrwaisYALW14} carried out a comprehensive study on the dark side of the parked domains to understand monetization, its scope, and magnitude. They controlled the start and end node of the monetization process, using an infiltration method, to send crawling traffic among specified start and end nodes with the monetization entities in between. Their analysis of one thousand seed redirection chains proved the presence of threats, such as click fraud, traffic spam, and traffic stealing in the monetization process. In addition, the results of this study showed that the revenue of the responsible parties for illicit monetization activities is as high as 40\% of the total revenues. 

\begin{table*}[t]
\begin{center}
\caption{Summary of the research works that have studied DNS privacy landscape.}
\label{tab:SumPrivacy}
\begin{tabular}{l|l|l|l}
\Xhline{2\arrayrulewidth}
\textbf{Work} & \textbf{Method}  & \textbf{strengths}    &\textbf{Weaknesses} \\
\Xhline{2\arrayrulewidth}
\cite{HerrmannFLF14} & EncDNS & Lightweight/Low-latency & Unresolveable internal domains \\ \hline
\cite{LuT10} & PPDNS & Improved privacy & Complex/DNS modification \\ \hline
\cite{MohaisenGR17} & DLV-Aware/PP-DLV & Improved privacy & MITM attacks/record modification \\
\cite{ZhaoHS07} & Range query & Simple/flexible & Bandwidth/response latency \\ \hline
\cite{ZhaoHS07-1} & PIR & Improved bandwidth  & Protocol modification\\ \hline
\cite{ZhuHHWMS15} & T-DNS & Improved privacy/security & Response latency \\ \hline
\cite{KrishnanM10} & No DNS prefetching & Inference attacks  & Inaccurate TTL/Unstable profiles\\ \hline
\Xhline{2\arrayrulewidth}
\end{tabular}
\end{center}
\end{table*}

Szurdi \etal~\cite{SzurdiKCSFK14} have conducted a large-scale measurement study to understand the impact of typosquatting on the monetization strategies of attackers among less known domains. Their analysis showed that large proportion (95.0\%) of the typosquatting domains are targeting less popular websites for domain monetization techniques \eg parked ads. Furthermore, Weaver \etal~\cite{WeaverKP11} have investigated the error traffic monetization practices of ISPs through analyzing 66,000 \textit{Netalyzr} session traces. Their analysis showed that ISPs were clearly involving on error traffic monetization by rerouting traffic to ad servers. 

Vissers \etal~\cite{VissersJN15} have investigated the ecosystem of parked domains from security view point to understand the security risks of parked domains on users. Their analysis showed that parked domains expose users into a set of threats, \eg malware, scam, and inappropriate content. Finally, they have presented a RF-based parked domain detection system through analyzing a set of generic features, including HTML features, HTTP archive features, frame Features, and domain name features. 

\BfPara{Discussion} Although domain parking is technically a legitimate business, it has been mixed up with malicious practices such as click fraud, malware, scam, malicious contents, \etc~\cite{VissersJN15, AlrwaisYALW14}. Dishonest parking services offer high revenue for responsible parties, which may compromise the benefits of legitimate advertisers and traffic buyers, by manipulating the traffic and redirecting them to rogue ad servers~\cite{AlrwaisYALW14}. Moreover, domain monetization is prevalent among both popular domains and less known domains~\cite{SzurdiKCSFK14}. In addition, ISPs actively involve traffic monetization activities~\cite{WeaverKP11}. 

\subsection{Privacy Leakage} \label{sec:Privacy}

A range of works have been conducted to address privacy-related issues of DNS \cite{Castillo-PerezG09, HerrmannFLF14, KrishnanM10, LuT10, PaxsonCJRSSSTVW13, MohaisenGR17, ZhuHHWMS15, LiuHW16, MohaisenR17, MohaisenKR16}. For example, Herrmann \etal~\cite{HerrmannFLF14} have explored the emerging threat of third-party DNS resolvers, \eg  Google Public DNS and OpenDNS to online privacy and introduced a lightweight privacy-preserving name resolution service called \textit{EncDNS}. EncDNS is designed based on the encapsulation of encrypted messages in standards-compliant DNS messages. EncDNS is compatible with existing popular DNS resolvers and offers low-latency DNS resolution.    
Furthermore, Shulman~\cite{Shulman14} have explored the existing proposals on DNS privacy that suggested encryption of DNS requests as a solution. The results demonstrated that a straightforward application of encryption alone may not provide the desired DNS privacy protection. 

Krishnan \etal~\cite{KrishnanM10} have explored the privacy implications of \textit{DNS prefetching}. Their analysis showed that an adversary can abuse the context inserted to a resolver's cache through prefetching to launch disclosure attacks, \eg reconstructing searched terms. Thus, they suggested that DNS prefetching should be turned off by default.

Zhu \etal~\cite{ZhuHHWMS15} have presented a connection-oriented DNS, called \textit{T-DNS}, to address the privacy and security issues of the connectionless DNS. T-DNS takes advantage of both TCP and TLS: while TCP protects the server against amplification attacks, TLS defends against eavesdroppers to the RDNS resolvers.
In addition, Reddy \etal~\cite{TDLS} have explored the protection of the privacy-sensitive information of DNS queries and responses. They presented a protection mechanism, called the \textit{Datagram Transport Layer Security (DTLS)}, for DNS exchange. DTLS counters passive listening and active attacks. The proposed mechanism reduces the DTLS round trips and the handshake size as well.

Zhao \etal~\cite{ZhaoHS07} have investigated the privacy disclosure of DNS queries and proposed a privacy-preserving query scheme, called \textit{Range Query}, which reduces privacy disclosure by concealing the actual queries using noisy traffic. However, Castillo-Perez and Garc{\'{\i}}a{-}Alfaro~\cite{Castillo-PerezG09} demonstrated that the privacy ensured by noisy traffic is not only difficult to analyze, but also does introduce undesired latency and bandwidth consumption. Castillo-Perez and Garc{\'{\i}}a{-}Alfaro~\cite{Castillo-PerezG09} have evaluated two DNS privacy-preserving proposals, the \textit{Range Query}~\cite{ZhaoHS07} and the \textit{Privacy Information Retrieval (PIR)} schemes~\cite{ZhaoHS07-1}. They demonstrated that both of these approaches are not desired. The first approach increases the latency and the bandwidth  during the execution and resolution of queries. Although the second approach is designed to address the limitations of the first approach, \eg bandwidth consumption, its functionality requires major modification of the DNS protocol, and relies on the use of DNSSEC.

Lu and Tsudik~\cite{LuT10} explored DNS privacy leaks during domain resolution and presented Privacy-Preserving DNS (PPDNS) that mitigates the privacy issues in DNS. PPDNS takes advantage of both distributed hash tables (DHTs) and computational private information retrieval (cPIR). While DHT is an alternative naming infrastructure that provides name resolution query privacy, cPIR reduces communication overhead. However, Federrath \etal~\cite{FederrathFHP11} argue that although PPDNS provides high level of privacy, it would not be adopted in the near future due to its computational complexity and requirement for completely different DNS infrastructure. 


DNS blocking of unintended queries is among the suggested solutions to improve the DNS privacy. For example, Appelbaum and Muffett~\cite{rfc7686} have studied Tor's privacy and suggested to block .onion names at stub, recursive, and authoritative resolvers to improve it's privacy. Mohaisen and Ren~\cite{MohaisenR17} have investigated the leakage of .onion at the A and J DNS root nodes over a longitudinal period of time and have found that .onion leakage is common and persistent at DNS infrastructure. Furthermore, Mohaisen \etal~\cite{MohaisenKR16} have studied the impact of blocking of unintended queries under different adversarial settings on the DNS privacy. The results highlighted that partial blocking at stub resolver would negatively affect the DNS privacy under certain adversary models; however, blocking the queries at the recursive would result in favorable privacy outcomes.

\BfPara{Discussion} Full and proper deployment of DNSSEC relies on cooperation between domain owners, name server owners, registries, and ISPs. Thus, various methods have been proposed to improve the privacy of DNS which are listed in~\autoref{tab:SumPrivacy} along with their strengths and weaknesses.  It should be noted that DNSSEC does not necessarily protect privacy. 
Although DNS over TLS protects against eavesdropping, it is not clear what  of privacy it offers. In addition, existing approaches require modifications to the DNS protocols~\cite{ZhaoHS07}. Privacy issues have been overlooked by DNS security efforts (such as DNSSEC) and are thus likely to propagate into future versions of DNS~\cite{LuT10}, requiring further attention.

\begin{table}
\begin{center}
\caption{Topical classification threats addressed in the literature, with sample work. MD=Malicious Domains, CP=Cache Poisoning,PH=Phishing, MA=Manipulation, AD=Amplification/DoS, DG=Domain Generation, B=Botnet, DS=Domain Squatting, M=Monetization, PL=Privacy Leakage.}
\label{tab:ThreatsCont.}

\begin{tabular}{l|l|l|l|l|l|l|l|l|l|l}
\Xhline{2\arrayrulewidth}
\textbf{Work}	& \rotatebox{90}{{\bf MD}} 	&	\rotatebox{90}{\textbf{CP}}	&	\rotatebox{90}{\textbf{PH}}	&	\rotatebox{90}{\textbf{MA}}	&	\rotatebox{90}{\textbf{AD}}	&	\rotatebox{90}{\textbf{DG}}	&	\rotatebox{90}{\textbf{B}}	&	\rotatebox{90}{\textbf{DS}}	&	\rotatebox{90}{\textbf{M}}	&	\rotatebox{90}{\textbf{PL}}	\\ 
\Xhline{2\arrayrulewidth}
\cite{ZouZPPLL16}	&		&		&		&		&		&		&		&		&		&	\cmark	\\	
\cite{AizuddinANNAA17}	&		&		&		&		&	\cmark	&		&		&		&		&		\\	\hline
\cite{ballaniF16}	&		&		&		&		&	\cmark	&		&		&		&		&		\\	\hline
\cite{HerzbergS14}	&		&		&		&		&	\cmark	&		&		&		&		&		\\	\hline
\cite{MacFarlandSK15}	&		&		&		&		&	\cmark	&		&		&		&		&		\\	\hline

\cite{PerdisciCG12}	&		&		&		&		&	\cmark	&		&		&		&		&		\\	\hline
\cite{TruongC16}	&		&		&		&		&	\cmark	&	\cmark	&	\cmark	&		&		&		\\	\hline
\cite{VermaHHHRKF16}	&		&		&		&	\cmark	&	\cmark	&		&		&		&		&		\\	\hline
 \cite{ShulmanW14}	&		&	\cmark	&		&		&		&		&		&		&		&		\\	\hline

\cite{AntonakakisDLPLB10}	&		&	\cmark	&		&		&		&		&		&		&		&		\\	\hline
\cite{ChenMP15}	&		&	\cmark	&		&		&		&		&		&		&		&		\\	\hline

 \cite{PerdisciALL09}	&		&	\cmark	&		&		&		&		&		&		&		&		\\	\hline
\cite{WuDZW15}	&		&	\cmark	&		&		&		&		&		&		&		&		\\	\hline

\cite{ChenAPNDL14}	&		&	\cmark	&		&		&		&		&		&		&		&		\\	\hline
\cite{HaoW17}	&		&	\cmark	&		&		&		&		&		&		&		&		\\	\hline

\cite{HerzbergS12}	&		&		&		&	\cmark	&		&		&		&		&		&		\\	\hline
\cite{jinYGY2015}	&	\cmark	&		&		&		&		&		&	\cmark	&		&		&		\\	\hline

\cite{JiangLLLDW12}	&	\cmark	&		&		&		&	\cmark	&		&	\cmark	&		&	\cmark	&		\\	\hline

 \cite{PlohmannYKBG16}	&		&		&		&		&		&	\cmark	&	\cmark	&		&		&		\\	\hline
 \cite{XuBSY13}	&		&		&		&		&	\cmark	&		&	\cmark	&		&		&		\\	\hline

 \cite{WeaverKNP11}	&		&		&		&	\cmark	&		&		&		&		&		&		\\	\hline
\cite{SchompCRA14}	&		&		&		&	\cmark	&		&		&		&		&		&		\\	\hline
 \cite{KuhrerHBRH15}	&	\cmark	&		&		&	\cmark	&		&		&		&		&		&		\\	\hline
\cite{PearceJLEFWP17}	&		&		&		&	\cmark	&		&		&		&		&		&		\\	\hline
\cite{BilgeKKB11}	&	\cmark	&		&	\cmark	&		&	\cmark	&		&	\cmark	&		&		&		\\	\hline
\cite{DagonPLL08}	&		&		&	\cmark	&		&		&		&		&		&		&		\\	\hline
 \cite{KintisMLCGPNA17}	&		&		&	\cmark	&		&		&		&		&	\cmark	&	\cmark	&		\\	\hline

\cite{AntonakakisPDLF10}	&	\cmark	&		&		&		&		&		&		&		&		&		\\	\hline
\cite{AntonakakisPLVD11}	&	\cmark	&		&	\cmark	&		&	\cmark	&		&	\cmark	&		&		&		\\	\hline
\cite{BilgeSBKK14}	&	\cmark	&		&		&		&		&		&		&		&		&		\\	\hline
\cite{CanaliCVK11}	&	\cmark	&		&		&		&		&		&		&		&		&		\\	\hline

\cite{FelegyhaziKP10}	&	\cmark	&		&		&		&		&		&		&		&		&		\\	\hline
\cite{GaoYCPGJD13}	&	\cmark	&		&		&		&		&		&		&		&		&		\\	\hline
\cite{HaoKMPF16}	&	\cmark	&		&		&		&		&		&		&		&		&		\\	\hline
\cite{HaoTPFKGH13}	&	\cmark	&		&		&		&		&	\cmark	&		&		&		&		\\	\hline

\cite{KhalilGNY18}	&	\cmark	&		&		&		&		&		&		&		&		&		\\	\hline
\cite{LuoTZSLNM15}	&	\cmark	&		&		&		&		&		&		&		&		&		\\	\hline

\cite{MaSSV09}	&	\cmark	&		&		&		&		&		&		&		&		&		\\	\hline

\cite{HaoFP11}	&	\cmark	&		&		&		&		&		&		&		&		&		\\	\hline
\cite{LiuLDWLD17}	&		&		&		&		&		&	\cmark	&		&	\cmark	&		&		\\	\hline
\cite{SzurdiKCSFK14}	&		&		&		&		&		&		&		&	\cmark	&	\cmark	&		\\	\hline
\cite{YadavRRR10}	&		&		&		&		&	\cmark	&	\cmark	&	\cmark	&		&		&		\\	\hline

 \cite{KhanHLK15}	&		&		&		&		&		&		&		&	\cmark	&		&		\\ \hline	
\cite{VissersBGJN17}	&		&		&		&		&		&		&		&	\cmark	&		&		\\	\hline
\cite{AlrwaisYALW14}	&		&		&		&		&		&		&		&		&	\cmark	&		\\	\hline
\cite{WeaverKP11}	&		&		&		&	\cmark	&		&		&		&		&	\cmark	&		\\	\hline

\cite{VissersJN15}	&		&		&		&		&		&		&		&	\cmark	&	\cmark	&		\\	\hline

\cite{HerrmannFLF14}	&		&		&		&		&		&		&		&		&		&	\cmark	\\	\hline

\Xhline{2\arrayrulewidth}

\end{tabular}
\end{center}
\end{table}

\begin{table}
\begin{center}
\caption{Topical classification of the DNS threats addressed in the literature, cont. of Table~\ref{tab:ThreatsCont.}.}
\label{tab:ThreatsCont2.}

\begin{tabular}{l|l|l|l|l|l|l|l|l|l|l}
\Xhline{2\arrayrulewidth}
\textbf{Work}	& \rotatebox{90}{{\bf MD}} 	&	\rotatebox{90}{\textbf{CP}}	&	\rotatebox{90}{\textbf{PH}}	&	\rotatebox{90}{\textbf{MA}}	&	\rotatebox{90}{\textbf{AD}}	&	\rotatebox{90}{\textbf{DG}}	&	\rotatebox{90}{\textbf{B}}	&	\rotatebox{90}{\textbf{DS}}	&	\rotatebox{90}{\textbf{M}}	&	\rotatebox{90}{\textbf{PL}}	\\ 
\Xhline{2\arrayrulewidth}
\cite{LuT10}	&		&		&		&		&		&		&		&		&		&	\cmark	\\	\hline

\cite{MohaisenGR17}	&		&		&		&		&		&		&		&		&		&	\cmark	\\	\hline

\cite{Tajalizadehkhoob17a}	&		&		&	\cmark	&		&		&		&		&		&		&		\\	\hline

\cite{LiAXYW13}	&		&		&		&		&	\cmark	&		&		&		&		&		\\	\hline
\cite{RahbariniaPA15}	&		&		&		&		&		&		&		&		&		&		\\	\hline
\cite{ShulmanW15}	&		&	\cmark	&		&		&		&	\cmark	&		&		&		&		\\	\hline

\cite{HerzbergS13}	&		&	\cmark	&		&		&		&		&		&		&		&		\\	\hline

\cite{JonesFPWA16}	&		&		&		&	\cmark	&		&		&		&		&		&		\\	\hline
\cite{ZhuHHWMS15}	&		&		&		&	\cmark	&		&		&		&		&		&		\\	\hline
\cite{KrishnanM10}	&		&		&		&		&		&		&		&		&		&	\cmark	\\	\hline
\cite{Castillo-PerezG09}	&		&		&		&		&		&		&		&		&		&	\cmark	\\	\hline
\cite{PaxsonCJRSSSTVW13}	&		&		&		&		&		&		&		&		&		&	\cmark	\\	\hline																			
\cite{MohaisenR17}	&		&		&		&		&		&		&		&		&		&	\cmark	\\	\hline
\cite{MohaisenKR16}	&		&		&		&		&		&		&		&		&		&	\cmark	\\	\hline
\cite{BenshoofRBH16}	&		&		&		&		&	\cmark	&		&		&		&		&		\\	\hline
\cite{DuYLDZ16}	&	\cmark	&		&		&		&		&		&		&		&		&		\\	\hline
\cite{LeverWNDMA16}	&	\cmark	&		&		&		&		&		&		&		&		&		\\	\hline
\cite{YuWLZ12}	&		&	\cmark	&		&		&		&		&		&		&		&		\\	\hline

\Xhline{2\arrayrulewidth}
\end{tabular}
\end{center}
\end{table}

\section{DNS Research Methods}\label{sec:Methods}

\begin{figure}[t]
\begin{center}
\includegraphics[width=0.4\textwidth]{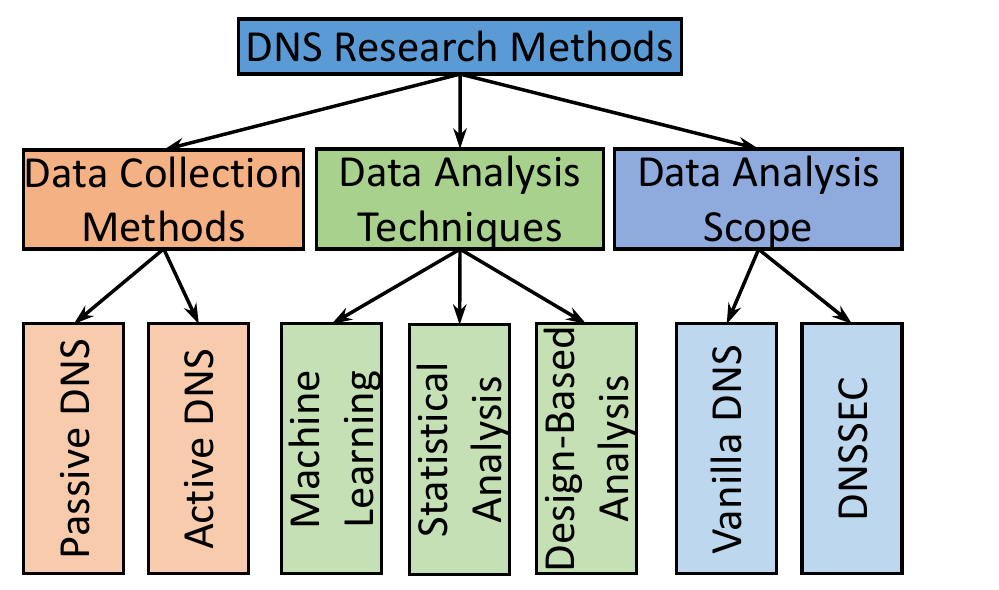}
\caption{DNS Research Methods. We have categorized DNS research method into data collection methods, data analysis techniques, and analysis scope.}
\label{DNSResearchMethods} 
\end{center}
\end{figure}

This section is devoted to investigating various data analysis methods, utilized to detect, model, and mitigate the aforementioned DNS threats. As it is outlined in \autoref{DNSResearchMethods}, the conducted research works will be categorized under data collection methods, data analysis techniques, and the scope of the analysis. This categorization will help us to better understand the advantages and disadvantages of different methodologies in the DNS security and privacy landscape, which will be an asset for future investigations. The topical classification of the DNS research methods, with sample works, is presented in \autoref{tab:ResearchMethods}.

\subsection{Data Collection Methods}

This section will describe two main DNS data collection methods, the passive DNS (PDNS) data and active DNS (ADNS) data, and the associated works.  

\subsubsection{Passive DNS} 
\begin{figure}[t]
\begin{center}
\includegraphics[width=0.45\textwidth]{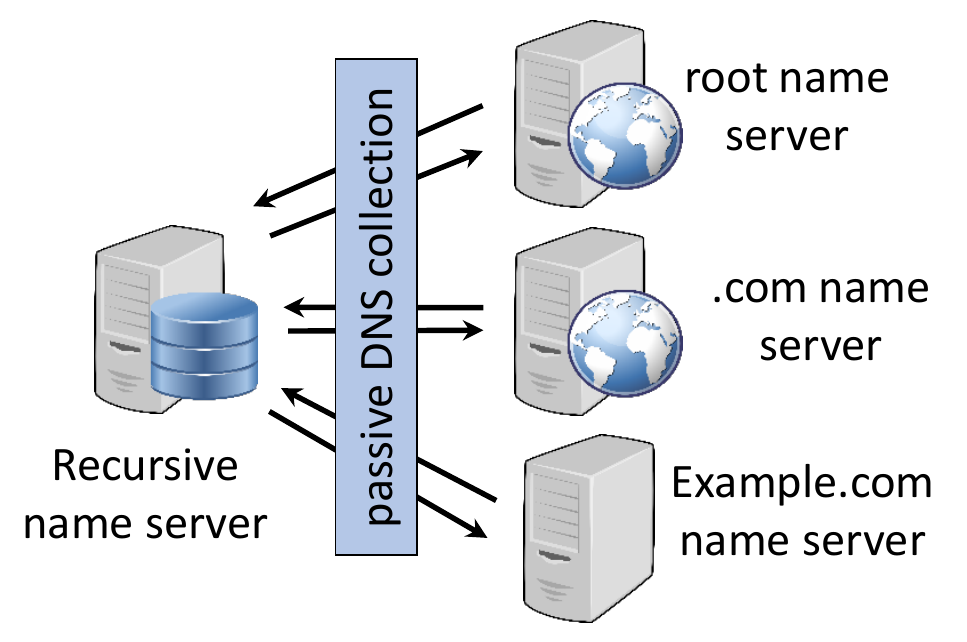}
\caption{High-level architecture of passive DNS collection system. In passive DNS collection, the DNS requests and responses are gathered from DNS providers across the internet as they happen.}
\label{fig:PassiveCollection}    
\end{center}
\end{figure}

Florian Weimer \cite{FlorianWeimer} has invented ``Passive DNS'' or ``Passive DNS replication'' technology to opportunistically reconstruct a partial view of the data available in the global DNS into a central database for further investigations. End-user's interactions can be monitored as they happen using passive monitoring~\cite{spring2012impact}. High-level architecture of passive DNS measurement systems is shown in~\autoref{fig:PassiveCollection}. Passive DNS databases based on valuable information they collect, have been considered as an invaluable asset of cybersecurity researchers to combat a wide range of threats such as malware, botnets, and malicious actors~\cite{AlrwaisLMWWQBM17, AntonakakisPLVD11, CallahanAR13, Tajalizadehkhoob17a, KountourasKLCND16, TruongC16, PerdisciCG12, LiAXYW13, BilgeSBKK14, KintisMLCGPNA17,  KhanHLK15, HaoW17, GaoYCPGJD13}. 

For example, Callahan \etal~\cite{CallahanAR13} have analyzed 200 million DNS queries collected through passive monitoring of DNS traffic of 90 home residential neighborhood network in the U.S.~\cite{casezone} in order to investigate the evolution and the behavior of the DNS servers and clients in modern DNS systems. Moreover, Gao \etal~\cite{GaoYCPGJD13} have analyzed 26 billion DNS query-response pairs collected by Security Information Exchange (SIE), now part of DNSDB~\cite{DNSDB}, from more than 600 recursive DNS resolvers distributed over North America and Europe, to empirically reexamine the performance and the operational characteristics of the DNS infrastructure. In addition, they found that temporal correlation analysis of the collected passive DNS queries can be used to detect malicious domain names. 

Hao and Wang~\cite{HaoW17} have studied 1.2 billion trace logs of outgoing DNS queries collected at local DNS servers at the College of William and Mary (WM) and the University of Delaware (UD) to understand the impact of one-time-use domain names on the performance of DNS caching. In addition, Alrwais \etal~\cite{AlrwaisLMWWQBM17} have analyzed 1.5 TB of passive DNS data collected by monitoring 25 snapshots of the entire IPv4 address space to systematically study the trend of BulletProof Hosting (BPH) services and potential detection and mitigation methods. 

Liu \etal~\cite{LiuLDWLD17} have queried passive DNS data from Farsight Security passive DNS database, DNSDB~\cite{DNSDB}, and the biggest public passive DNS system in China~\cite{NetLab} to detect malicious domains. Moreover, Tajalizadehkhoob \etal~\cite{Tajalizadehkhoob17a} have analyzed 1,259 shared hosting providers extracted from DNSDB\cite{DNSDB} to explore the distribution of web security features and software patching practices in shared hosting providers.

\begin{table*}[t]
\begin{center}
\small
\caption{Summary of the data collection methods used in research works. Here TB is Terra Byte, M is Million, B is Billion, and LDNS donates local DNS servers. {$^{(1)}$}WM and {$^{(2)}$} are the Collage of William and Mary and the University of Delaware, respectively.}
\label{tab:SumDataCollection}
\scalebox{0.85}{
\begin{tabular}{l|l|l|c|c|c|c|c|c|c|c|c|c|c|c|l|l}
\Xhline{2\arrayrulewidth}

\multirow{2}{*}{\textbf{Work}} & \multirow{2}{*} {\textbf{Method}} & \multirow{2}{*} {\textbf{Volume}} & \multicolumn{12}{c|}{\textbf{Query Types}} & \multirow{2}{*}{\textbf{Data Scale}} & \multirow{2}{*}{\textbf{Source}} \\ \cline{4-15}
                            &  &  & \rotatebox{90}{A}  & \rotatebox{90}{AAAA} & \rotatebox{90}{MX} & \rotatebox{90}{NS} & \rotatebox{90}{TXT} & \rotatebox{90}{PTR} & \rotatebox{90}{SRV} & \rotatebox{90}{SOA} & \rotatebox{90}{CNAME} & \rotatebox{90}{DS} & \rotatebox{90}{DNSKEY} & \rotatebox{90}{NSEC} & & \\
                            \Xhline{2\arrayrulewidth}
                            
\multirow{3}{*}{\cite{HaoW17}}
 & PDNS &  1.01B & \cmark  & \cmark &  & \cmark & \cmark & \cmark & \cmark & \cmark &  &  &  &  & 2015-06 to 2015-07 & LDNS at WM{$^{(1)}$}\\ \cline{2-17} 
 & PDNS &  0.19B & \cmark  & \cmark & & \cmark & \cmark & \cmark & \cmark & \cmark &  &  &  &  & 2015-12 & LDNS at UD{$^{(2)}$}\\ \hline

\multirow{2}{*}{\cite{KintisMLCGPNA17}}
 & PDNS & 13.1B  &  &  &  &  &  &  & &  &  &  &  & & 2011-01 to 2015-10& Largest ISP in the U.S. \\ \cline{2-17}
 & ADNS & 455B & \cmark  & \cmark& \cmark & \cmark & \cmark &  & & \cmark & \cmark &  &  &  & 2015-10 to 2016-08 & com/name/net/org/biz/Alexa 1M  \\ \hline               

\cite{GaoYCPGJD13} & PDNS &  26B & \cmark  & \cmark & \cmark &  &  & \cmark &  &  &  &  &  &  & & 600 RDNS \\ \hline

\cite{AlrwaisLMWWQBM17} & PDNS & 1.7TB & \cmark & \cmark & \cmark & \cmark &  &  &  &  &  &  &  &  &  2015-01 to 2016-08 & 25 snapshots of IPv4 list \\ \hline

\cite{CallahanAR13} & PDNS &  200M  & \cmark  & \cmark &  &  &  & \cmark &  &  &  &  &  &  & 2011-01 to 2012-03 & 90 home residential network\\ \hline

\cite{KountourasKLCND16} & ADNS & Source list & \cmark  & \cmark & \cmark & \cmark & \cmark &  &  & \cmark & \cmark &  &  &  & Daily since 2016 & com/name/net/org/biz/Alexa 1M  \\ \hline

\cite{VanJSP16} & ADNS & Source list & \cmark  & \cmark & \cmark & \cmark & \cmark &  &  & \cmark &  \cmark& \cmark & \cmark & \cmark  & Daily since 2015 & com/net/org/info/mobi/name/biz/asia/aero \\ 
\Xhline{2\arrayrulewidth}
\end{tabular}}
\end{center}
\end{table*}

\begin{table}[t]
\begin{center}
\small
\caption{List of common query types and their description. }
\label{tab:querytypes}
\begin{tabular}{l|l}
\Xhline{2\arrayrulewidth}
\textbf{Query Type}  & \textbf{Definition} \\
\Xhline{2\arrayrulewidth}
A	&	IPv4 address\\		
AAAA&	IPv6 address \\
MX	&	Mail exchanger record\\
NS	&	Authoritative name server\\
TXT	&	Arbitrary text strings\\
PTR	&	Pointer (IP address/hostname)\\
SRV	&	Service (service/hostname)\\
SOA	&	Start of Authority\\
CNAME&  Canonical Name (Alias/canonical)\\
DS	&	Delegation of Signing\\
DNSKEY	&	DNSSEC public key\\
NSEC	&	Next SECure (No record/two points)\\
\Xhline{2\arrayrulewidth}
\end{tabular}
\end{center}
\end{table}

Furthermore, Khan \etal~\cite{KhanHLK15} have utilized passive DNS data to study the negative impact of the typosquatting on the users through the time it takes to find the intended website. Kintis \etal~\cite{KintisMLCGPNA17} have analyzed 13.1 billion passive DNS RRs collected from the largest ISP in the U.S., to understand how combosquatting is used by miscreants for malicious purposes. 

Bilge \etal~\cite{BilgeSBKK14} have analyzed the passive DNS data obtained by SIE~\cite{ISC} to detect and block malicious domain names in real time. In addition, Li \etal~\cite{LiAXYW13} have investigated the topological relation of hosts and malicious websites using the same PDNS API, SIE~\cite{ISC}. Moreover, Perdisci \etal~\cite{PerdisciCG12} have utilized the SIE PDNS traffic to detect and track malicious domain-flux networks. Similarly, Truong \etal~\cite{TruongC16} have analyzed passive DNS traffic for detection of domain-flux botnets within a monitored LAN network.

Although passive DNS data is widely used to study and address to various vulnerabilities of DNS, \eg malware, botnets, domain squatting, \etc~\cite{GaoYCPGJD13, HaoW17,KhanHLK15,TruongC16}, the collection of such data is costly, challenging, and needs restrictive legal agreements~\cite{KountourasKLCND16}.

\subsubsection{Active DNS} 
\begin{figure}[t]
\begin{center}
\includegraphics[width=0.46\textwidth]{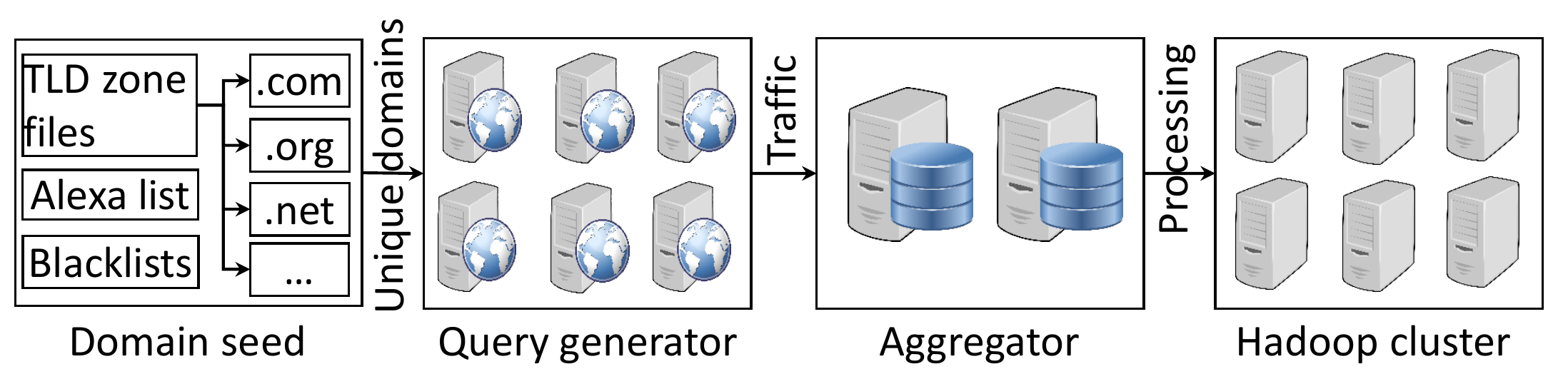}
\caption{High-level architecture of a typical active DNS collection system. Active DNS collection is composed of four main steps including gathering domain seeds, generating queries, collecting responses, and storing/processing the data in Hadoop cluster.}
\label{ActiveDNS_arch} 
\end{center}
\end{figure}

The concept of active DNS measurement have been presented to tackle the potential barriers of DNS security research. Active DNS measurement cannot be a replacement for passive DNS measurement; however, it offers several advantages in comparison to its counterpart~\cite{KountourasKLCND16}. For example, Internet activities of real users would not fluctuate active measurements. In addition, active DNS measurements could be largely available with least privacy concerns, while it offers an order of magnitude more domain names and IP addresses. Therefore, researchers have developed and used active DNS measurement systems to address DNS security threats~\cite{ KhanHLK15, JonesFPWA16, KountourasKLCND16, VanJSP16, KhalilGNY18, KintisMLCGPNA17, ChungR0CLMMW17, jonker2016measuring}. Generally, active DNS collection systems are composed of four main components: domain seeds, query generator, collection point, and storage. The list of the domains to be queried are gathered in domain seeds. These lists are requested using query generator and the responses are collected at the collection point and then stored in Hadoop cluster for further analysis. High-level architecture of active DNS collection system is shown in~\autoref{ActiveDNS_arch}. 

Kountouras \etal~\cite{KountourasKLCND16} have presented an active DNS measurement called Thales, which reliably queries, collects, and distills active DNS datasets. A set of large-scale seed domain list including public blacklists~\cite{BHDNS, DNSBHP, ZeusT}, the Alexa list~\cite{AlexaTop}, the Common Crawl dataset~\cite{CommonCrawl}, the domain feed from an undisclosed security vendor, and the zone files for the TLDs consisting of .com, .net, .biz and .org constitute the list of seed domains. Moreover, Van Rijswijk-Deij \etal~\cite{VanJSP16} have presented OpenINTEL, an active DNS measurement system that conducts daily measurement for all of the domains in .com, .net, and .org TLDs. OpenINTEL reliably queries and collects over 50\% of the global DNS data based on daily active measurements and since February 2015.

Moreover, Khan \etal~\cite{KhanHLK15} have actively crawled 13.5k suspected domains to study the impact of typosquatting events on both typical users and legitimate websites. Kintis \etal~\cite{KintisMLCGPNA17} have analyzed 455 billion DNS RRs collected daily through the active DNS project~\cite{ActiveDNSProject, KountourasKLCND16} from November 2015 to August 2016, to investigate how attackers utilize combosquatting to conduct variety of malicious activities including trademark infringement, malware, monetization, and phishing. Moreover, Khalil \etal~\cite{KhalilGNY18} have conducted an in-depth analysis of active DNS dataset provided by Thales~\cite{ActiveDNSProject, KountourasKLCND16} to design an association-based scheme for detection of malicious domains with high accuracy and coverage.

Chung \etal~\cite{ChungR0CLMMW17} have collected all DNSSEC-enabled domain names, under the .com, .org, and .net TLDs through OpenINTEL active DNS measurement system~\cite{VanJSP16}, to study the adoption and management of DNSSEC in both authoritative name servers and DNS resolvers. Moreover, Toorn \etal ~\cite{ToornRGS18} have studied the detection of Snowshoe spams through active DNS measurement and using machine learning techniques. 

\BfPara{Discussion} Passive monitoring is better suited as a technology to determine what goes wrong with a website or web application after problems have occurred. Passive DNS, though, is rare, difficult to obtain, and often comes with restrictive legal clauses (i.e., non disclosure agreements). At the same time, laws and regulations against personal identifiable information, the significant financial cost of the passive collection, and storage infrastructure are some of the reasons that make passive DNS difficult. In addition, passive DNS datasets are limited in scope and time. On the other hand, although active DNS datasets offer several unique characteristics, the amount of offered information would be limited. It should be noticed that active DNS measurement highly relies on the domain seeds. In addition, issued DNS queries based on the limited set of hosts yields a localized DNS data collection. In other words, the collected DNS data is limited to the locations of those hosts, thus all IPs associated with a specific given name might not be identified. Furthermore, DNS queries are made by data collectors not by individual users, which makes active DNS data less suitable as an approach to detect malicious domains based on user level features of DNS data, such as user query patterns. The summary of the collected dataset in the literature is in~\autoref{tab:SumDataCollection} along with the query types description in~\autoref{tab:querytypes}.  

\subsection{Data Analysis Methods}
Cybersecurity researchers have employed variety of data analysis techniques to effectively address to DNS threats. For example, the application of machine learning algorithms are widely used \cite{AlrwaisLMWWQBM17, AntonakakisDLPLB10, LiuLDWLD17}. This section introduces some of the prevalent data analysis techniques that have been used in the cybersecurity community, such as machine learning algorithms \cite{AlrwaisLMWWQBM17, AntonakakisDLPLB10, AntonakakisPDLF10, HaoTPFKGH13, HaoW17, KhalilGNY18, LuoTZSLNM15, MaSSV09, RadwanH17, RahbariniaPA15, ShulmanW15, VissersJN15} and association analysis \cite{YadavRRR10, HockK16, KhalilGNY18, GomezNA17, GaoYCPGJD13}.

\subsubsection{Machine Learning Algorithm} 
Antonakakis \etal~\cite{AntonakakisDLPLB10} have designed an Machine Learning (ML)-based cache poisoning detection system, called Anax, which detects malicious changes in cached DNS records in real time with 91.9\% of detection accuracy rate. Similarly, Hao and Wang~\cite{HaoW17} have trained Decision Tree (DT) and Random Forest (RF) machine learning algorithms using syntactical features of one-time-use domain names, \eg domain name string, the length of query name \etc to detect and expel one-time-use domain names from DNS cache in an attempt to increase the performance of DNS caching.

Alrwais \etal~\cite{AlrwaisLMWWQBM17} have utilized two machine learning algorithms, namely SVM and RF, to build a detetction system to identify malicious network blocks using features extracted from the trends of the BPH services. Khalil \etal~\cite{KhalilGNY18} have designed a malicious domain detection system using RF model and based on the representative features of dedicated IPs, \eg the number of fully qualified domain names, the number of second level domains in its /24 IP block \etc

\begin{table*}[t]
\begin{center}
\caption{Summary of the machine learning methods used in the literature. Here AR is Accuracy Rate, while FPR represents False Positive Rate. The abbreviations of ML algorithms are described in~\autoref{tab:abbreviations} }
\label{tab:SumML}
\begin{tabular}{l|l|c|c|c|c|c|c|l|l}
\Xhline{2\arrayrulewidth}
\multirow{2}{*}{\textbf{Work}} & \multirow{2}{*} {\textbf{Application}} & \multicolumn{6}{c|}{\textbf{ML algorithm} } & \multirow{2}{*}{\textbf{AR}} & \multirow{2}{*}{\textbf{FPR}}  \\ \cline{3-8}
                            &  & DT & RF & NB & KNN & SVM & MLP & &  \\
\Xhline{2\arrayrulewidth}
\cite{AntonakakisDLPLB10} & Cache poisoning & \cmark &  & \cmark & \cmark & \cmark & \cmark & 91.9\% & 0.6\%  \\ \hline
\cite{HaoW17} & Cache poisoning & \cmark & \cmark &  &  &  &   & 88.0\% & 1.0\%  \\ \hline
\cite{KhalilGNY18} & Malicious domain & \cmark & \cmark &  &  & \cmark &   & 99.0\% & 1.0\%  \\ \hline
\cite{VissersJN15} & Parked domains &  & \cmark &  &  &  &   & 98.7\% & 0.5\%  \\ \hline
\cite{AlrwaisLMWWQBM17} &  Malicious network  &  & \cmark & &  & \cmark &  & 97.1\% & 1.6\%  \\ \hline
\cite{liXFZ13} & Phishing &  &  &  &  & \cmark &    & 95.5\% & 3.5\%  \\ 
\Xhline{2\arrayrulewidth}
\end{tabular}
\end{center}
\end{table*}

Liu \etal~\cite{LiuLDWLD17} have utilized a set of machine learning models including NN, RF, LR, SVM, and NB to automatically identify shadowed domains with 98.5\% accuracy rate. In addition, Vissers \etal~\cite{VissersJN15} have proposed a parked domain detection system based on RF algorithm that is trained over a set of the generic and robust features of parked domains. Moreover, Li \etal \cite{liXFZ13} have used TSVM algorithm instead of classical SVM to improve the detection accuracy for phishing webpages. 

\subsubsection{Association Analysis} The strong associations of domains with known malicious domains can be utilized to detect malicious domain names. For example, Khalil \etal~\cite{KhalilGNY18} have designed an association-based scheme for detection of malicious domains with high accuracy and coverage. Furthermore, Gao \etal~\cite{GaoYCPGJD13} have utilized the temporal correlation analysis of DNS queries to identify a wide range of correlated malicious domain groups, \eg phishing, spam, and DGA-generated domains based on the related known malicious anchor domains. Yadav \etal~\cite{YadavRRR10} have utilized statistical measures such as Kullback-Leibler divergence, Jaccard index,and Levenshtein for domain-flux botnet detection. Gomez \etal~\cite{GomezNA17} have studied the application of visualization for understanding the DNS-based network threat analysis.

\BfPara{Discussion} Despite numerous advantages of machine learning approaches, there are still risks and limitations of using them in operation. The foremost challenge is the acquisition and labeling of relevant data from representative vantage points to maximize insights. Even if the the data is collected correctly, capturing DNS traffic results in a very large amount of data to analyze, which would be expensive in term of computation and storage. In addition, the performance of machine learning algorithms is contingent upon their structure and learning algorithms. It should be noticed that selecting improper structure or learning algorithms might result in poor results; thus, it is mandatory to try different algorithms for each problem. Furthermore, the training phase of the algorithm would be a time-consuming process, even if the dataset is small, requiring training heuristics.  

\begin{table}[t]
\begin{center}
\caption{Topical classification of the DNS research methods addressed in the literature, with sample work}
\label{tab:ResearchMethods}

\begin{tabular}{l|l|l|l|l|l|l|l|l|l}
\Xhline{2\arrayrulewidth}

Work	&	\rotatebox{90}{\textbf{PDNS}}  	&	\rotatebox{90}{\textbf{ADNS}}	&\rotatebox{90}{\textbf{Analysis}}	&\rotatebox{90}{\textbf{Scope}}	 & Work	&	\rotatebox{90}{\textbf{PDNS}}  	&	\rotatebox{90}{\textbf{ADNS}}	&\rotatebox{90}{\textbf{Analysis}}	&\rotatebox{90}{\textbf{Scope}}\\ 
\Xhline{2\arrayrulewidth}

\cite{AdrichemBLWWFK15}	&		&		&		&	\cmark	 	&	 \cite{JonesFPWA16}	&		&	\cmark	&		&		\\ \hline
\cite{AlrwaisLMWWQBM17}	&	\cmark	&		&	\cmark	&		 	& \cite{KhalilGNY18}	&		&	\cmark	&	\cmark	&		\\ \hline
 \cite{AntonakakisDLPLB10}	&		&		&	\cmark	&		 	&	 \cite{KhanHLK15}	&	\cmark	&	\cmark	&		&		\\ \hline
 \cite{AntonakakisPDLF10}	&		&		&	\cmark	&		 	&	 \cite{KintisMLCGPNA17}	&	\cmark	&	\cmark	&		&		\\ \hline
\cite{AntonakakisPLVD11}	&	\cmark	&		&		&		 	&	 \cite{KountourasKLCND16}	&	\cmark	&	\cmark	&		&		\\ \hline
 \cite{BilgeSBKK14}	&	\cmark	&		&		&		 	& \cite{LuoTZSLNM15}	&		&		&	\cmark	&		\\ \hline
\cite{CallahanAR13}	&	\cmark	&		&		&		 	&	 \cite{MohaisenGR17}	&		&		&		&	\cmark	\\ \hline
\cite{CanaliCVK11}	&		&		&	\cmark	&		 	&	 \cite{PaxsonCJRSSSTVW13}	&		&	\cmark	&		&		\\ \hline
\cite{ChenAPNDL14}	&	\cmark	&		&		&	\cmark	 	&	 \cite{PerdisciALL09}	&		&		&		&	\cmark	\\ \hline
 \cite{ChungR0CLMMW17}	&	\cmark	&	\cmark	&		&	\cmark	 	&	 \cite{PerdisciCG12}	&	\cmark	&		&		&		\\ \hline
 \cite{ChungRCLMMW17}	&		&		&		&	\cmark	 	& \cite{RadwanH17}	&		&		&	\cmark	&		\\ \hline
\cite{DuYLDZ16}	&		&	\cmark	&		&		 	&  \cite{RahbariniaPA15}	&		&	\cmark	&	\cmark	&		\\ \hline
\cite{fofack2013modeling}	&	\cmark	&		&		&			&	\cite{ShulmanW14}	&		&		&		&	\cmark	\\ \hline
\cite{GaoYCPGJD13}	&	\cmark	&		&		&		 	&	\cite{ShulmanW15}	&		&		&	\cmark	&		\\ \hline
 \cite{HaoTPFKGH13}	&		&		&	\cmark	&		 	&	\cite{ShulmanW17}	&		&		&		&	\cmark	\\ \hline
\cite{HaoW17}	&	\cmark	&		&	\cmark	&		 	&	\cite{Tajalizadehkhoob17a}	&	\cmark	&		&		&		\\ \hline
\cite{HerzbergS13}	&		&		&		&	\cmark	 	&	 \cite{TruongC16}	&	\cmark	&		&		&		\\ \hline
\cite{HerzbergS12}	&		&		&		&	\cmark	 	&	\cite{VanJSP16}	&		&	\cmark	&		&		\\ \hline
\cite{HerzbergS14}	&		&		&		&	\cmark	 	&	 \cite{VissersJN15}	&		&		&	\cmark	&		\\ \hline
\cite{JalalzaiSI15}	&		&		&		&	\cmark	 	&	&   &    &     & \\
\Xhline{2\arrayrulewidth}

\end{tabular}
\end{center}
\end{table}

\subsection{Data Analysis Scope}

The original design of the DNS did not consider any security details; instead, it was designed to be a scalable distributed system.  As the Internet has grown, malicious actors have found weaknesses, mainly due to lack of DNS records verification, in the DNS system that allows them to launch variety of attacks, \eg phishing, malware, \etc. Thus DNSSEC was created to secure the DNS infrastructure. In this section, we will categorize the research works from security point of view into vanilla DNS~\cite{SchompCRA14, DagonPLL08, XuBSY13, JacksonBBSB09, SchompAR14, DagonAVJL08, ChenMP15} and DNSSEC~\cite{AdrichemBLWWFK15, BabuP18,BauM10, ChungR0CLMMW17, ChungRCLMMW17, HerzbergS13, HerzbergS12, HerzbergS14, JalalzaiSI15}. This will help us to understand the advantages, \eg improved security and disadvantages, \eg overheads of the deployment of the DNSSEC. 

\subsubsection{Vanilla DNS} Although DNS is the largest distributed system, it is vulnerable to multiple threats, such as DNS cache poisoning attacks, amplification attacks, \etc. For example, Schomp \etal~\cite{SchompCRA14} have demonstrated that user-side DNS infrastructure is vulnerable to record injection threats. Moreover, Xu \etal~\cite{XuBSY13} have demonstrated that attackers can severely abuse DNS-based stealthy C\&C channel to efficiently hide malicious DNS activities. In addition, Schomp \etal ~\cite{SchompAR14} have showed that shared DNS resolvers are vulnerable to various forms of attack~\cite{HerzbergS13, SchompCRA14}, \eg fraudulent record injection.

In order to address the security problems of the DNS infrastructure, which are mainly due to the lack of integrity and authenticity of DNS records, the utilization of DNSSEC have been suggested~\cite{HerzbergS12,JalalzaiSI15}.

\subsubsection{DNS Security Extension} 
\begin{figure}[t]
\begin{center}
\includegraphics[width=0.35\textwidth, angle =90]{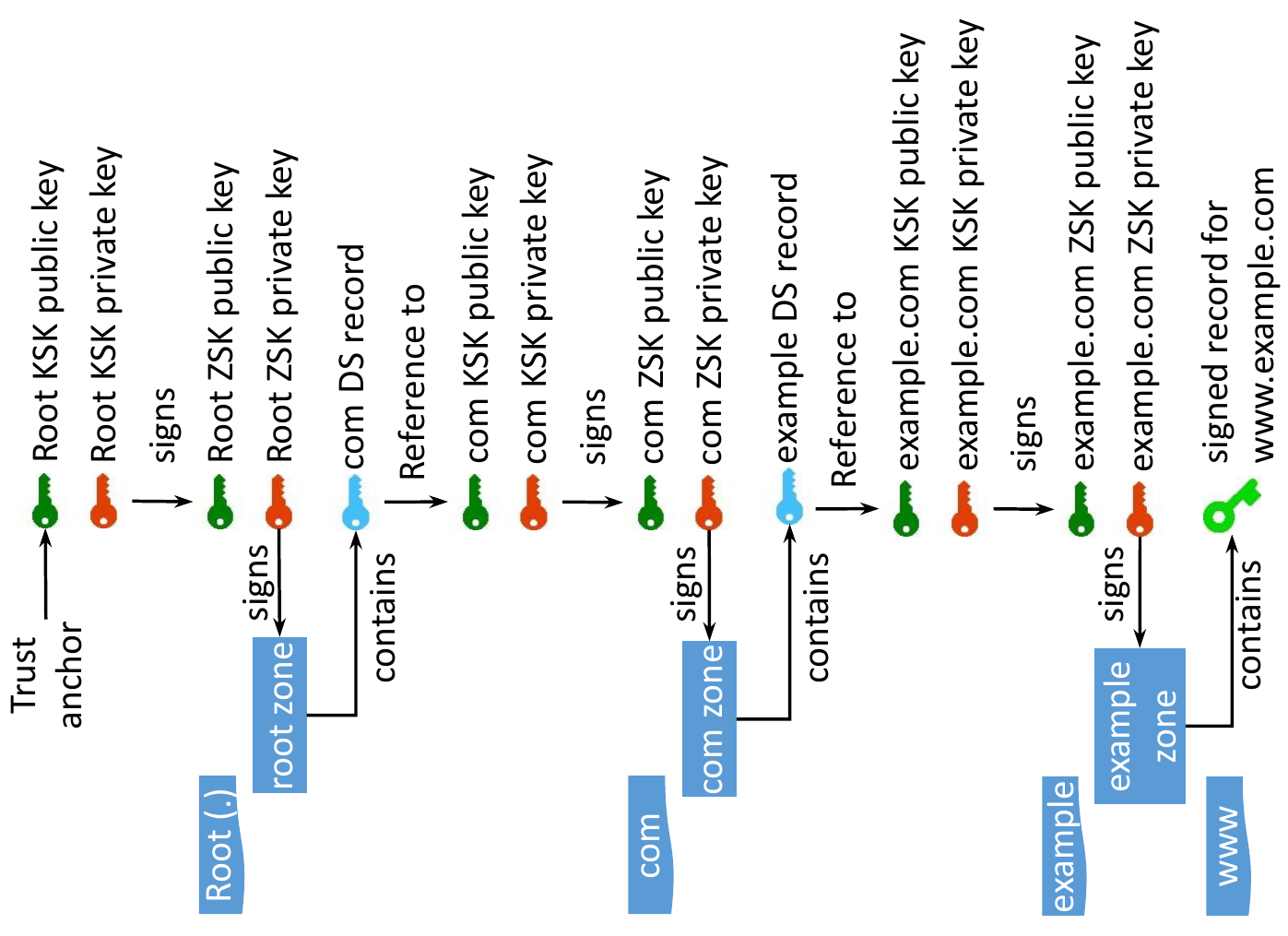}
\caption{An overview of DNSSEC chain of trust. DNSSEC utilizes public and private key pairs to verify the integrity and authenticity of DNS records based on a chain-of-trust. The verification process of DNS record begins at the root and continues until the record in question is authenticated.}
\label{fig:DNSSECChain}    
\end{center}
\end{figure}

DNSSEC allows clients to verify the integrity and authenticity of DNS records based on a chain-of-trust. Each zone in DNSSEC is composed of two public and private key pairs, including Key Signing Key (KSK) and Zone Signing Key (ZSK). Public Key Infrastructure (PKI) of the DNSSEC is rooted at the KSK of the DNS root zone. Thus, the validation process of DNS response begins at the root and continues until the record is authenticated. An overview of the DNSSEC chain of trust is shown in \autoref{fig:DNSSECChain}.



Chung \etal~\cite{ChungR0CLMMW17} have conducted a measurement study on deployment and management of DNSSEC in both authoritative name servers and DNS resolvers. They found that only a tiny fraction ($\approx$1\%) of all domains in three top TLDs publish DNSKEY, including .com, .net, and .org. Additionally, their results showed that over 30\% signed domains fail to upload DS records, highlighting the failure of authoritative name servers in uploading records for a majority of their domain names. Towards the client side of DNSSEC, Lian \etal~\cite{LianRSS13} have studied the capability of the resolvers and end-users to achieve DNSSEC authentication, and concluded that  DNSSEC increases the failure rate of end-to-end resolutions.  

\begin{figure}[t]
\begin{center}
   \includegraphics[width=0.5\textwidth]{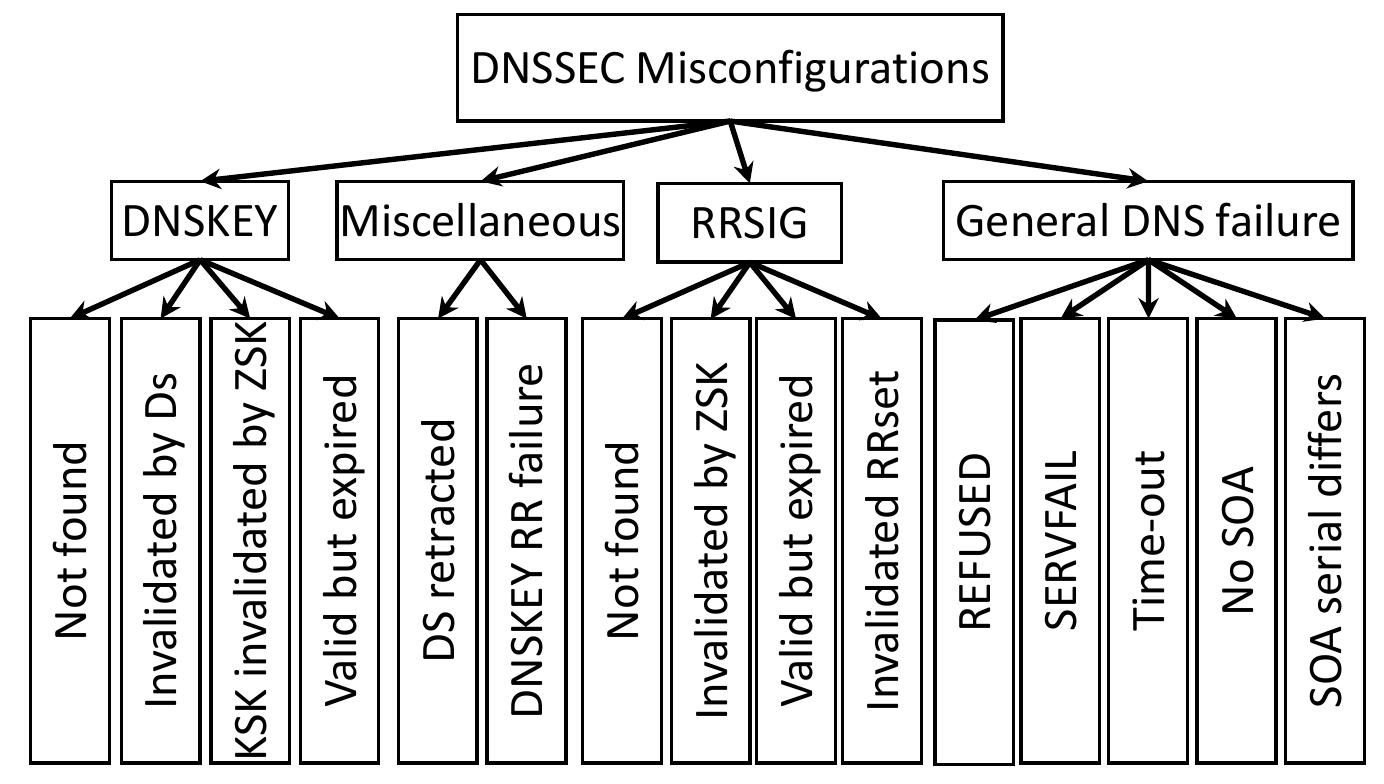}
   \caption{Overview of DNSSEC misconfigurations. In general, DNSSEC misconfigurations can be categorized in four groups, including DNSKEY, RRSIG, general DNS failure, and miscellaneous.}
   \label{fig:DNSSECMisconf} 
\end{center}
\end{figure}

To categorize the potential root causes of the DNSSEC misconfigurations, Adrichem \etal~\cite{AdrichemBLWWFK15}, have conducted a measurement study to access the impact of misconfigurations on the reachability of zone's network. Their analysis showed that only 7.93\% of gathered domain names from .bg, .br, .co, and .se zones attempted to implement DNSSEC, and over 4\% were misconfigured. Furthermore, they observed that the DNSSEC-aware resolvers were unable to reach a large fraction ($\approx$73.86\%) of misconfigured domains. The impact of DNSSEC misconfigurations on DNS query requests has also been addressed by Deccio \etal~\cite{DeccioSKM11}, and in general, DNSSEC misconfigurations can be broadly classified into four categories: DNSKEY, RRSIG, general DNS failure, and miscellaneous.~\autoref{fig:DNSSECMisconf} shows the categories and the subcategories of DNSSEC misconfigurations~\cite{AdrichemBLWWFK15,DeccioSKM11}. 

DNS operators have a critical role in the deployment of DNSSEC, as they are responsible for the maintenance of DNSKEY and RRSIG records. Chung \etal~\cite{ChungRCLMMW17} have studied the impact of DNS operators, \eg registrar, owner, or third-party DNS operator, on DNSSEC deployment. They have observed that many popular registrars fail to support DNSSEC, and as such, there are only 3 mutual registrars among the top 25 popular, and the top 25 fully deployed registrars.

\BfPara{Discussion} Correct deployment of DNSSEC relies on end-to-end establishment of RRSIGs, a cryptographically valid DNSKEY, and DS records. However, a small fraction of all domains ($\approx$1\%) in the top three TLDs (.com, .net, and .org) publish DNSKEY~\cite{ChungR0CLMMW17}. Moreover, incomplete or incorrect deployment of DNSSEC may lead to attacks such as domain hijacking. The threat does not completely mitigate even after full deployment. An adaptive attacker may still succeed in the attack by using a fake name server, that enables off-path traffic analysis and a covert channel~\cite{HerzbergS13}. The hazards of DNSSEC misconfiguration are not limited to the failures; misconfigured domains are always at risk of being unreachable from a DNSSEC-aware resolver.

\section{DNS Entities Scopes}\label{sec:Entities}

\begin{figure}[t]
\begin{center}
   \includegraphics[width=0.3\textwidth]{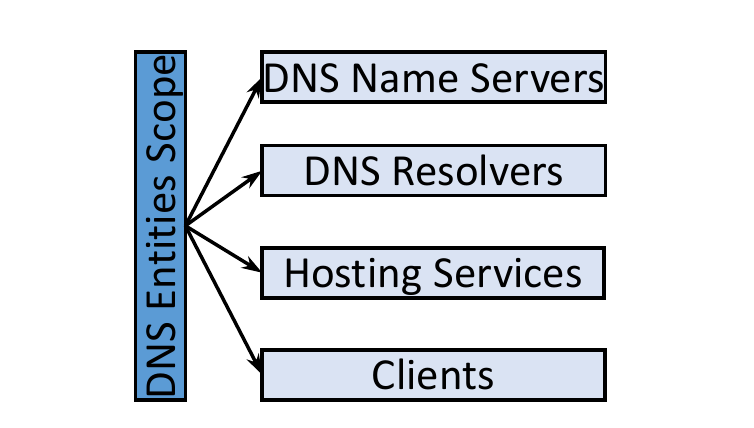}
   \caption{DNS Entities Scope. We have categorized research works based on the main scope of the entities into DNS name servers, open DNS resolvers, hosting services, and clients.}
   \label{fig:DNSEntitiesScope} 
\end{center}
\end{figure}

DNS infrastructure comprises of different entities, including DNS name servers, DNS resolvers, hosting providers, and clients. Each entity is designed to serve a specific task. DNS infrastructure's performance is contingent upon the collaboration and synchronization of these components. In this section, we survey the prior work to provide deeper insights about the impact of different DNS entities on the overall system. We provide an outline of this section in \autoref{fig:DNSEntitiesScope}, and in \autoref{tab:DNS_Entity}, we present the topical classification of the DNS entities as addressed in the literature, with sample works. 

\subsection{DNS Name Servers}

A DNS server provides name resolution of names into IP addresses, and vice versa. Extensive work has been published to shed light on the name servers' landscape \cite{HaoWSS15, AntonakakisPLVD11, ChenWRZ07, ShangW06, KangJKRM16, VissersBGJN17, ShulmanW14, LiangJDLW13, GaoYCPGJD13, MohaisenBL14}. 


\subsubsection{Authoritative Name Server} An authoritative name server is responsible for answering queries related to a particular domain name in a zone. The server response contains actual DNS records of the queried domain name, such as A, CNAME, PTR, \etc, which highlight the importance of the server's performance and security. List of the common query types are reported in~\autoref{tab:querytypes}. 
Hao \etal~\cite{HaoWSS15} have investigated the characteristics of DNS, \eg performance, availability, life-cycle and deployment patterns. In their work, they analyzed three major types of DNS servers namely private, upstream, and hybrid authoritative DNS servers, and showed that a majority of websites host the authoritative DNS servers in upstream services for better performance. 

Liang \etal~\cite{LiangJDLW13} studied the impact of top level DNS servers' uneven distribution on the end-user latency. They observed that despite good performance of top level DNS severs, the quality of the service was unfair in different continents including  Africa and South America. In those regions, the performance was 3--6 times worse than Europe and North America. Furthermore, they made two useful observations:
\begin{enumerate*}
    \item Almost all the root servers in Europe and North America work efficiently. 
    \item Only F, J, and L roots show low query latency in other continents. 
\end{enumerate*}

Malicious domains can be identified by the number of name servers that a domain name interacts with over time. For instance, Felegyhazi \etal~\cite{FelegyhaziKP10} have utilized the inherent characteristics of domain registrations and their appearance in DNS zone files to build a proactive malicious domain detection system. Mohaisen \etal~\cite{MohaisenBL14} presented \textit{Name Server Switching Footprint (NSSF)}, a feature modality and a system that analyze and detect the domain names with suspicious name server switching behavior. Additionally, in the same work, they have designed a time series prediction model to predict the number of name servers that a domain is likely to interact with.  

Due to the critical role of the authoritative name servers in DNS infrastructure, they have been analyzed from a security point of view as well. For example, Vissers \etal~\cite{VissersBGJN17} have investigated the security practices of authoritative name servers, and observed that a large number of domain names ($\approx$1.28M) are under the risk of DoS attacks due to weak security practices of few ($\approx$7,214) authoritative name servers. They also suggested employment of DNSSEC as a solution to enhance the security of DNS. Furthermore, Chung \etal~\cite{ChungR0CLMMW17} have conducted a measurement study on deployment and management of DNSSEC in both authoritative name servers and resolvers. Their results showed that more than 30\% of signed domains failed to upload DS records, and some authoritative name servers failed to upload DS records for nearly all of their domains.  Such practices, make DNSSEC ineffective and provide no practical security.   

\subsubsection{Recursive DNS server} A recursive DNS server is responsible to query other name servers to resolve a domain name. While recursive DNS resolver is significantly important in the function of DNS, miscreants have exploited it to launch various attacks, \eg DNS amplification attack, cache poisoning, \etc Accordingly, researchers have explored various aspects of recursive DNS resolvers including, caching performance, DNS amplification attack, malware infection, \etc~\cite{ChenWRZ07,AntonakakisPLVD11, ChenAPNDL14,VermaHHHRKF16,HaoW17}.  Chen \etal~\cite{ChenAPNDL14} have investigated the impact of disposable domains on the caching behavior of the DNS resolvers. Their analysis showed that as the prevalence of the disposable domains increases, it is likely that the DNS cache begins to fill up with RRs, that are unlikely to be reused. Hao and Wang~\cite{HaoW17} have studied the negative impact of one-time-use domain names on the performance of DNS caching. They argue that removal or not insertion of such RRs into the cache can prevent a waste of the DNS cache resources. 

Shang \etal~\cite{ShangW06} explored the relationships across domain names to improve the cache hit rate of local DNS servers. In addition, Chen \etal~\cite{ChenWRZ07} have presented a dynamic lease scheme that keeps track of local DNS name servers to provide stronger cache consistency and more reliable services. Perdisci \etal~\cite{PerdisciALL09} have proposed a novel method based on WSEC DNS to mitigate recursive DNS resolvers’ cache poisoning attack. 

Ager \etal ~\cite{AgerMSU10} explored the impact of DNS resolvers latency and the DNS cache contents on the performance of local DNS and open DNS resolvers, \eg GoogleDNS~\cite{GoogleDNS} and OpenDNS~\cite{OpenDNS}. They further observed that unlike third-party DNS resolvers, local DNS resolvers redirect clients towards the content available within the ISP. Kuhrer \etal \cite{KuhrerHBRH15} conducted a large-scale study based on empirical data that is collected over one year, to investigate the landscape of DNS resolvers. Their analysis revealed that millions of these DNS resolvers deliberately manipulate resolutions for malicious purposes. 

Verma \etal~\cite{VermaHHHRKF16} have utilized query rate sharing property of DNS resolvers to build a DNS DDoS mitigation system that calculates global DNS query rate to make mitigation decisions locally. Moreover, Ballani \etal~\cite{ballaniF16} have analyzed the caching behavior of DNS resolvers to defend against DNS DoS attacks. 

\BfPara{Discussion} Although top level DNS servers perform well, the quality of service is unfair around the globe \cite{LiangJDLW13}. Moreover, it is noteworthy that the vulnerability of a small number of authoritative name servers, potentially affects a large number of domains under their apex~\cite{VissersBGJN17}. Authoritative name servers are widely exploitable through outdated WHOIS email records of name server domains. Moreover, few authoritative name servers are responsible for incorrect deployment of DNSSEC in large number of domain names\cite{ChungR0CLMMW17}.

\subsection{Open DNS Resolver}
\begin{figure}[t]
        \centering
        \includegraphics[width=0.44\textwidth]{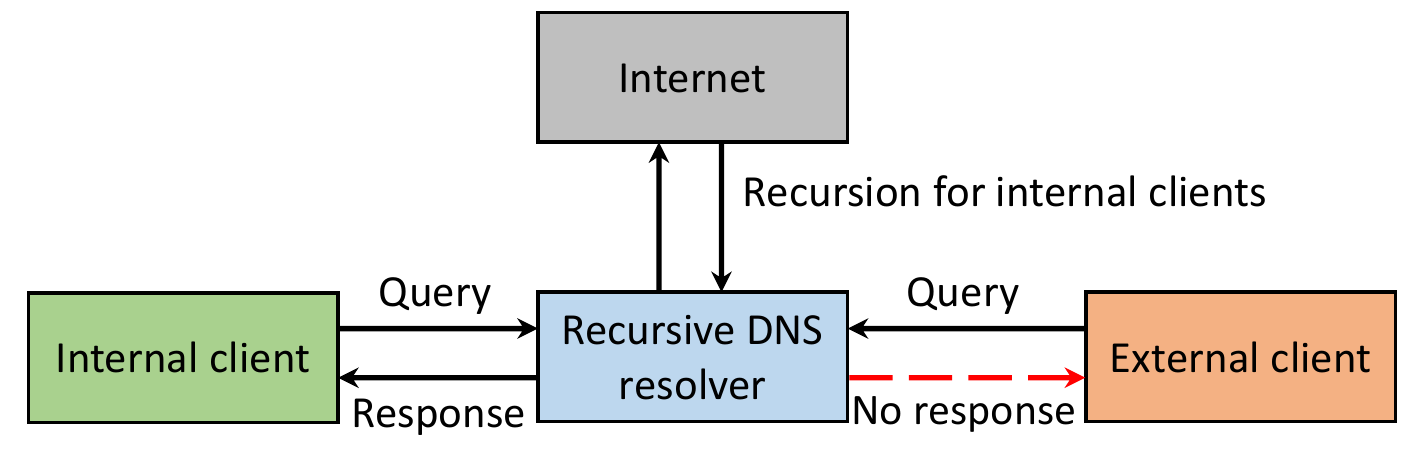}
        \centering
        \caption{Overview of recursive DNS resolver. Notice that RDNS resolver does not respond to queries of external clients.}
        \label{RDNS}
\end{figure}

\begin{figure}[t]
        \centering
        \includegraphics[width=0.44\textwidth]{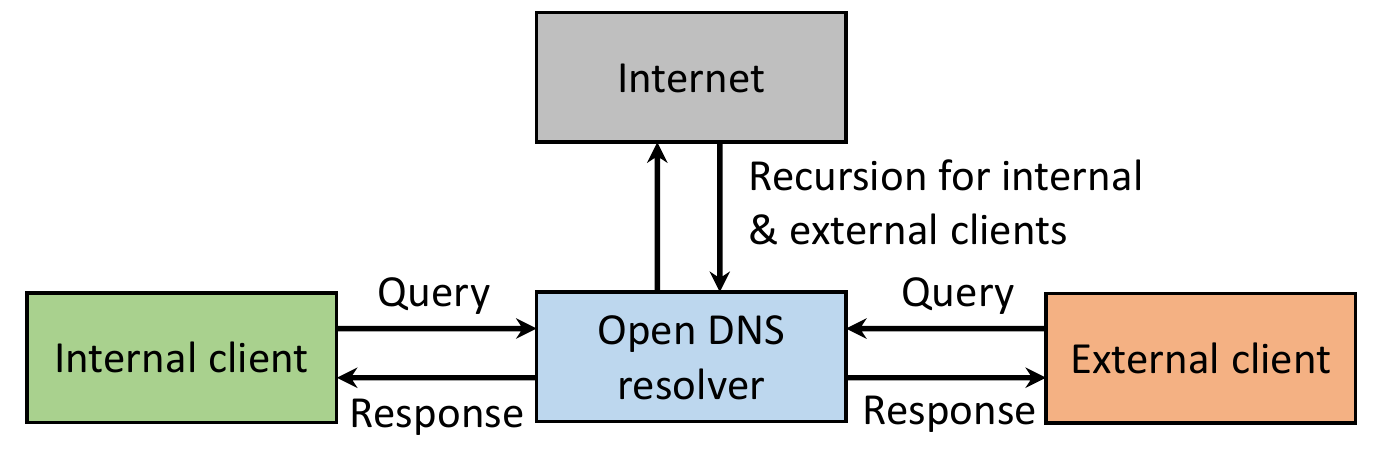}
        \centering
        \caption{Overview of open recursive DNS resolver. Notice that ORDNS resolver does respond to queries of both internal and external clients.}
        \label{ODNS}
\end{figure}

Contrary to the recursive DNS resolvers, that perform recursion for the internal clients only, an open DNS resolver resolves recursive DNS lookups for anyone over the Internet. As such, an open DNS resolvers is vulnerable to well known attacks, \eg DDoS on behalf of attackers. An overview of recursive DNS resolvers and open DNS resolvers is shown in~\autoref{RDNS} and ~\autoref{ODNS}, respectively. To this end, extensive research efforts have been devoted to study the impact of open resolvers on the performance and security of DNS infrastructure~\cite{AgerMSU10, AntonakakisDLPLB10, ballaniF16, GaoYCPGJD13, HerrmannFLF14, KuhrerHBRH15, HerzbergS12, PearceJLEFWP17, PerdisciALL09, SchompAR14, SchompCRA14, VermaHHHRKF16}.

Based on the aforementioned properties of open resolvers, they can be exploited by the attackers to conduct a series of attacks. Significant research has been conducted to understand the DNS resolvers landscape and its associated threats.

\subsubsection{Measurement on the DNS Protocol} Prior work on the analysis of DNS resolvers is mostly focused on a small subset of all resolvers. Therefore, it is unclear if the observations can be generalized on a wider scale to all the resolvers around the globe. For instance, Sisson \cite{sisson2010dns} analyzed open resolvers based on sampled scans that repeatedly query the same set of resolvers. Therefore, their work was limited to only a small fraction of all the open resolvers. Antonakakis \etal~\cite{AntonakakisDLPLB10} have analyzed a geographically diverse set of 300,000 open resolvers to measure the integrity of their responses. They observed that in general, the attackers point victims to rogue IP addresses. Furthermore, Jiang \etal \cite{JiangLLLDW12} analyzed the caching behavior of resolvers and identified an attack vector in DNS software that allows the extension of domains caching even after their removal from the upper DNS hierarchy.

\begin{table}[t]
\begin{center}
\small
\caption{Summary of the research works that measured the number of open DNS resolvers.}
\label{tab:SumORDNS}
\begin{tabular}{llll}
\Xhline{2\arrayrulewidth}
\textbf{Work}  & \textbf{Year} & \textbf{Scope} & \textbf{\# ORDNS}    \\
\Xhline{2\arrayrulewidth}
\multirow{2}{*}{\cite{sisson2010dns}} & 2010&  105.4M IPs  & 11.9M   \\ \cline{2-4}
  & 2010& 1.1M  IPs & 114.7K  \\ \hline
\cite{SchompCRA13} & 2013 &  IPv4 IPs  & 32M  \\ \hline
\multirow{2}{*}{\cite{Mauch13, openresolverproject}} & 2013 &  IPv4 IPs  & 32M \\ \cline{2-4}
 & 2017 &  IPv4 IPs  & 10.3M  \\ \hline
\multirow{2}{*}{\cite{shadowserver}} & 2013 & IPv4 IPs  & 11.9M  \\\cline{2-4} 
 & 2017 & IPv4 IPs  & 3.7M  \\ \hline

\multirow{2}{*}{\cite{WesselsM14}} & 2013 & IPv4 IPs  & 33.6M  \\\cline{2-4} 
 & 2014 & IPv4 IPs  & 29.2M  \\ \hline

\cite{takano2013measurement} & 2013 & IPv4 IPs & 25M \\ 

\Xhline{2\arrayrulewidth}
\end{tabular}
\end{center}
\end{table}

Schomp \etal \cite{SchompCRA13} randomly probed the IPv4 address space to enumerate DNS resolvers and distinguish between recursive DNS resolvers and DNS proxies. They also performed an in-depth analysis on the caching behavior of the resolvers. Gao \etal~\cite{GaoYCPGJD13} have analyzed a large set of DNS query-response pairs collected from over 600 recursive DNS resolvers. They observed that despite a great variation in the characteristics of the DNS traffic across networks, the behavior of resolvers within an organization remains similar.

\subsubsection{Internet-Wide Scanning} Durumeric \etal~\cite{DurumericWH13} proposed a high-speed application to run Internet-wide scans called {\em ZMap}. {\em ZMap} is an open-source network scanner designed to perform Internet-wide scans with capability of surveying the entire IPv4 address space within 45 minutes. The Open Resolver Project \cite{Mauch13, openresolverproject} actively investigates DNS servers world-wide and provides open resolver statistics on the web. The open resolver statistics are available online from March 2013 to January 2017, after which the scan was discontinued. Shadowserver \cite{shadowserver} is another organization that conducts surveys related to the Internet security including active measurements of open resolvers, with up-to-date scans. Furthermore, Takano \etal \cite{takano2013measurement} performed measurements based on responses for Internet-wide DNS software version requests. They have focused on DNS server software and their distribution in each regional Internet registry.

\subsubsection{Threats} Antonakakis \etal~\cite{AntonakakisDLPLB10} noticed that in general, the attackers point victims to rogue IP addresses through open DNS resolvers. Kuhrer \etal~\cite{KuhrerHBRH15} have analyzed the threats in open resolvers from two perspectives. Firstly, they scanned the DNS resolvers' landscape for changes in the course of time and categorized the resolvers based on their device type and the software version. Secondly, they measured the response authenticity of the resolvers from the users' point-of-view to find that a large number of resolvers intentionally manipulate DNS resolutions for malicious activities.


Schomp \etal~\cite{SchompCRA14} have measured the vulnerability of the user-side DNS infrastructure to record injection threats. They have found that many open DNS resolvers, that are vulnerable to record injection attacks, are used as an attack vector to target shared DNS infrastructure. Verma \etal~\cite{VermaHHHRKF16} presented a system to mitigate Amplified DNS DDoS (ADD) attacks, that relies on the fact that DNS resolvers share local DNS query rates that can be used to calculate the total query rates.


Ballani \etal~\cite{ballaniF16} have presented a defense scheme called stale cache, that uses caching behavior to prevent DNS DoS attacks.They have analyzed DNS traces under multiple DoS attack strategies and have found that the cost-benefit ratio of the proposed method, favors the deployment of stale cache. Furthermore, Herrmann \etal~\cite{HerrmannFLF14} have studied third party DNS resolvers and consequent threats to online privacy. They proposed a new privacy-preserving name resolution service that relies on the encapsulation of encrypted messages in standards-compliant DNS messages.  Hendriks \etal~\cite{HendriksSRP17} presented an active measurement system to find the list of open DNS resolvers on IPv6 in the wild, which could be potentially exploited in a DDoS attack. Moreover, Schomp \etal~\cite{SchompAR14} have studied the vulnerabilities of shared DNS resolvers and proposed an unorthodox method to tackle such threats.


\BfPara{Discussion} DNS resolvers can be easily discovered and used by the attackers for malicious activities, such as DoS attacks. Moreover, buggy implementations and large DNSSEC response filtering can lead to a high query latency\cite{LiangJDLW13}. Although extensive research has been conducted on open DNS resolvers, their evolution and distribution still require a clear demonstration. Some open questions that are worth addressing in this regard are: 
\begin{enumerate*}
    \item How is the distribution of open DNS resolvers in popular TLDs?
    \item  Are open resolvers consistent in answering various clients for the same query type? 
\end{enumerate*}

\begin{table}[t]
\begin{center}
\caption{Topical classification of the DNS research scopes addressed in the literature, with sample work; NS=Name Servers, R=Resolvers, HS=Hosting Services, and C=Clients.}
\label{tab:DNS_Entity}
\begin{tabular}{l|l|l|l|l|l|l|l|l|l}
\Xhline{2\arrayrulewidth}

Work	&	\rotatebox{90}{\textbf{NS}} 	&\rotatebox{90}{\textbf{R}}		&\rotatebox{90}{\textbf{HS}}		&\rotatebox{90}{\textbf{C}} & Work	&	\rotatebox{90}{\textbf{NS}} 	&\rotatebox{90}{\textbf{R}}		&\rotatebox{90}{\textbf{HS}}		&\rotatebox{90}{\textbf{C}}		\\ 
\Xhline{2\arrayrulewidth}

 \cite{AdrichemBLWWFK15} &  &  & \cmark & 	&	\cite{LiangJDLW13}	&	\cmark	&		&		&		\\ \hline
\cite{AgerMSU10}	&	\cmark	&	\cmark	&		&		&	\cite{LuoTZSLNM15}	&		&		&		&	\cmark	\\ \hline
\cite{AntonakakisDLPLB10} &  & \cmark &  &  	&	\cite{MaSSV09}	&		&		&	\cmark	&		\\ \hline
 \cite{AntonakakisPLVD11} & \cmark &  &  &  	&	 \cite{PanYC16}	&		&		&		&	\cmark	\\ \hline
\cite{ballaniF16} &  & \cmark &  &  	&	\cite{PearceJLEFWP17}	&		&	\cmark	&		&		\\ \hline
\cite{ChenWRZ07} & \cmark &  &  &  	&	\cite{PerdisciALL09}	&		&	\cmark	&		&		\\ \hline
\cite{ChungR0CLMMW17} & \cmark &  & \cmark &  	&	\cite{QinXWJK14}	&		&	\cmark	&		&		\\ \hline
\cite{GaoYCPGJD13} &  & \cmark &  &  	&	 \cite{SchompAR14}	&		&	\cmark	&		&		\\ \hline
\cite{GreschbachPRWF17} &  & \cmark &  &  	&	 \cite{SchompCRA14}	&		&	\cmark	&		&	\cmark	\\ \hline
\cite{HaoW17} & \cmark &  &  &  	&	\cite{SchompRA16}	&		&		&		&	\cmark	\\ \hline
\cite{HaoWSS15} & \cmark &  & \cmark &  	&	\cite{ShangW06}	&	\cmark	&		&		&		\\ \hline
\cite{HerrmannFLF14} &  & \cmark &  &  	&	\cite{ShulmanW14}	&	\cmark	&		&		&		\\ \hline
\cite{HerzbergS13} & \cmark & \cmark &  &  	&	\cite{ShulmanW15}	&		&	\cmark	&		&		\\ \hline
\cite{HerzbergS12} &  & \cmark &  &  	&	\cite{Tajalizadehkhoob17a}	&		&		&	\cmark	&		\\ \hline
\cite{HerzbergS14} &  &  & \cmark &  	&	\cite{TruongC16}	&	\cmark	&		&		&		\\ \hline
\cite{JiangLLLDW12} &  & \cmark &  &  	&	\cite{VermaHHHRKF16}	&		&	\cmark	&		&		\\ \hline
\cite{KhanHLK15} &  &  &  & \cmark 	&	\cite{VissersBGJN17}	&	\cmark	&		&		&		\\ \hline
\cite{KuhrerHBRH15} &  & \cmark &  & \cmark 	&	\cite{YuWLZ12}	&		&	\cmark	&		&		\\ \hline
\cite{LiAXYW13} &  &  & \cmark &  	& &&&&	 \\
\Xhline{2\arrayrulewidth}
\end{tabular}
\end{center}
\end{table}

\subsection{Hosting services}

The task for remedying compromised web resources is shared between hosting providers and webmasters. Shared hosting providers retain more control over configurations, which explains their association with the widespread abuse. 


In~\cite{AdrichemBLWWFK15}, Adrichem \etal have conducted measurements to identify and categorize potential causes of the DNSSEC misconfigurations based on the reachability of a zone's network. They have analyzed domains in six zones, including .bg, .br, .co, .com, .nl and .se, and noticed that a small number of hosting providers are responsible for DNSSEC misconfigured domains. They concluded that the misconfigured domains are at the risk of being unreachable from DNSSEC-aware resolver. DNSSEC allows clients to verify the integrity and authenticity of DNS records. However, DNSSEC has wittnessed a low deployment rate with only 0.6\% of .com, 0.8\% of .net, and 1\% .org  properly signed domains~\cite{ChungR0CLMMW17}. Chung \etal~\cite{ChungR0CLMMW17} have attempted to identify the impact of registrars on deployment of DNSSEC, since the registrars usually serve as DNS operators to their customers. A DNS operator can be a registrar, an owner, or a third-party operator, responsible for maintaining DNSKEY and RRSIG records. Based on the registrars' policy of uploading records, DNSSEC can be fully deployed (with DS record), or partially deployed (without DS record). 


Hao \etal~\cite{HaoWSS15} have conducted a measurement study to understand the authoritative DNS servers' deployment patterns in modern web services. They have explored several characteristics DNS servers' including, performance, availability, and life-cycle of servers. They have heuristically analyzed the Alexa's top 1-milion list to identify the authoritative DNS deployment patterns in web domains. Their results show that most of the emerging popular social websites host the authoritative DNSes in upstream services that provide performance advantages. Furthermore, they observed that backup and redundant deployment in hybrid patterns provide availability; revealing the growth of cloud providing DNS hosting services.  Herzberg \etal~\cite{HerzbergS14} have studied DNS-amplification DoS attacks, and have proposed a defense system that is compatible with common DNS servers configurations and DNSSEC. The have shown the efficiency and high performance of the presented DNS-authentication method in preventing DNS-based amplification DoS attacks. They have also predicted the adoption of their design, based on game theory algorithms, concluding that their design will sufficiently reduce DNS amplification DoS attacks. Additionally, their proposed method can be deployed as a cloud-based service, that can reduced cost and upgrade defenses for DNS servers.

\BfPara{Discussion} Cloud services are increasingly popular due to their affordability \cite{HaoWSS15}. However, they are also subject to DoS attacks, where clients are at risk due to poorly written programs. Clients of shared hosting providers share IP address with other clients, which means that blocking one IP results in blocking all the users. Therefore, the security of cloud services is critical and should be well established.

\subsection{Client}
Although, much  knowledge about DNS infrastructure has been derived from aggregate population of clients' behavior analysis, researchers have also emphasized on individual client's interaction with DNS ecosystem \cite{KhanHLK15, KuhrerHBRH15, LuoTZSLNM15, PanYC16, SchompCRA14, SchompRA16, GreschbachPRWF17}. For example, Schomp \etal~\cite{SchompRA16} have analyzed the behavior of individual DNS users in order to develop an analytical model of their interaction with the DNS ecosystem. They observed that different types of users behave uniquely with DNS. In addition, they have demonstrated that the combination of the Weibull and Pareto can successfully model the process of client query arrival. Finally, they observed the existence of a fairly stable and unique working set of names for each client.


Pan \etal~\cite{PanYC16} have proposed a novel client classification method based on client query entropy and global recursive DNS service architecture. By monitoring the query frequencies of the clients, they have validated the effectiveness of the proposed method on busy and long-tailed clients. They have found that 2.32\% clients can cover the most important web spiders, recursive servers, and well-known internet services. Greschbach \etal~\cite{GreschbachPRWF17} have investigated the impact of DNS traffic on Tor clients' vulnerability to correlation attacks and how DNS lookups can be utilized for information stealth.

Kuhrer \etal~\cite{KuhrerHBRH15} have highlighted the vulnerabilities in DNS resolvers by analyzing the changes in their landscape, and the response authenticity of the resolvers from the users' perspective. They observed that a large number of resolvers intentionally manipulate resolutions for malicious activities.

Khan \etal~\cite{KhanHLK15} have measured the harm caused to the users by quantitatively measuring the time wasted by typosquatting. They have introduced a new metric that offers several advantages such as empirical quantification of harm to the clients, identification of typosquatting domain names, and proportion of different typosquatting perpetrators. They have analyzed a large scale DNS dataset (active and passive), and noticed that the typosquatting events affect both the users and the websites. Their analysis confirms that typosquatting increases both the time taken by the client to find the intended website, and the latency between a typo and its correction.  


Luo \etal~\cite{LuoTZSLNM15} have analyzed ISP network traces to understand various approaches of the attackers towards the exploitation of DNS for malicious activities. They have taken a variety of syntactic as well as temporal patterns into account to propose a method that identifies different clusters of malicious domain names that lead to DNS failures. Based on their evolutionary learning framework, less suspicious clusters were removed while highly suspicious cases were preserved. Their proposed framework analyzes DNS failures on per-client basis and in practice, they have used their framework on a large-scale ISP network trace to find that over 97\% of the users with suspicious DNS activities can be detected with 81\% precision.

\BfPara{Discussion} Clients are usually the main target in cyber crimes, and they suffer from a wide variety of attacks including phishing attacks, DNS manipulation, cybersquatting, \etc Existing research in the literature has
analyzed the impact of threats on the clients, however, it remains an open challenge to extend the literature for user-intention-based anomaly detection method to identify anomalous DNS traffic.

\section{Conclusion}\label{sec:Conclusion}

In this paper, a comprehensive study have been conducted to review the ecosystem of DNS from various points of view. Therefore, a large number of peer reviewed and recently published research works have been surveyed and then potential vulnerabilities of DNS security and corresponding countermeasures have been summarized. In addition, we have looked into the research works from two different point of view, namely data analysis methods and DNS entities landscape. In DNS research methods, we have looked into the literature to understand the limitations as well as strengths of utilized data analysis methods which would offer valuable guideline to analyze the existing challenges in DNS ecosystem. The second, DNS entities scope, looks into the role of different entities in DNS infrastructure and points out common vulnerabilities and limitations of existing components of the DNS ecology. Furthermore, in each sub-section we have discussed the challenges and pointed out open research directions. 

\section{Acknowledgement}
This work is supported in part by National Research Foundation of South Korea under grant NRF-2016K1A1A2912757.


\section*{References}

\bibliographystyle{unsrt}
\bibliography{amin,conf}

\end{document}